\documentclass[sigconf,screen]{acmart} 

\usepackage{multirow,array,booktabs,siunitx}

\definecolor{ForestGreen}{RGB}{34,139,34}
\definecolor{Crimson}{RGB}{220,20,60}
\newcommand{\edit}[1]{\textcolor{black}{#1}}

\usepackage{colortbl,booktabs,wasysym}
\usepackage{makecell}
\usepackage{enumerate}
\usepackage{amsmath}
\usepackage{enumitem}
\definecolor{CB_lightCyan}{HTML}{99DDFF}
\definecolor{CB_darkCyan}{HTML}{66CCFF}
\definecolor{CB_pear}{HTML}{BBCC33}
\definecolor{CB_pink}{HTML}{E4CEE1}
\newcommand{\Exp}{\textcolor{ForestGreen}{\CIRCLE}}
\newcommand{\Imp}{\textcolor{orange!80!black}{\LEFTcircle}}
\newcommand{\Abs}{\textcolor{gray}{\Circle}}

\AtBeginDocument{%
  }

\copyrightyear{2026}
\acmYear{2026}
\setcopyright{cc}
\setcctype{by}
\acmConference[CHI '26]{Proceedings of the 2026 CHI Conference on Human Factors in Computing Systems}{April 13--17, 2026}{Barcelona, Spain}
\acmBooktitle{Proceedings of the 2026 CHI Conference on Human Factors in Computing Systems (CHI '26), April 13--17, 2026, Barcelona, Spain}
\acmPrice{}
\acmDOI{10.1145/3772318.3790476}
\acmISBN{979-8-4007-2278-3/2026/04}

\begin{document}

\title[The Language of Approval: Identifying the Drivers of Positive Feedback Online]{The Language of Approval:\\Identifying the Drivers of Positive Feedback Online}

\author{Agam Goyal}
\orcid{0009-0009-5989-2887}
\affiliation{%
  \institution{University of Illinois Urbana-Champaign}
  \city{Urbana}
  \state{Illinois}
  \country{USA}}
\email{agamg2@illinois.edu}

\author{Charlotte Lambert}
\orcid{0009-0002-8487-7485}
\affiliation{%
  \institution{University of Illinois Urbana-Champaign}
  \city{Urbana}
  \state{Illinois}
  \country{USA}}
\email{cjl8@illinois.edu}

\author{Eshwar Chandrasekharan}
\orcid{0000-0002-7473-1418}
\affiliation{%
  \institution{University of Illinois Urbana-Champaign}
  \city{Urbana}
  \state{Illinois}
  \country{USA}}
\email{eshwar@illinois.edu}

\renewcommand{\shortauthors}{Goyal et al.}

\begin{abstract}
Positive feedback via likes and awards is central to online governance, yet which attributes of users' posts elicit rewards---and how these vary across authors and communities---remains unclear. To examine this, we combine quasi-experimental causal inference with predictive modeling on 11M posts from 100 subreddits. We identify linguistic patterns and stylistic attributes causally linked to rewards, controlling for author reputation, timing, and community context. For example, overtly complicated language, tentative style, and toxicity reduce rewards. We use our set of curated features to train models that can detect highly-upvoted posts with high AUC. Our audit of community guidelines highlights a ``policy-practice gap''---most rules focus primarily on civility and formatting requirements, with little emphasis on the attributes identified to drive positive feedback. These results inform the design of community guidelines, support interfaces that teach users how to craft desirable contributions, and moderation workflows that emphasize positive reinforcement over purely punitive enforcement.
\end{abstract}




\begin{teaserfigure}
    \includegraphics[width=\textwidth]{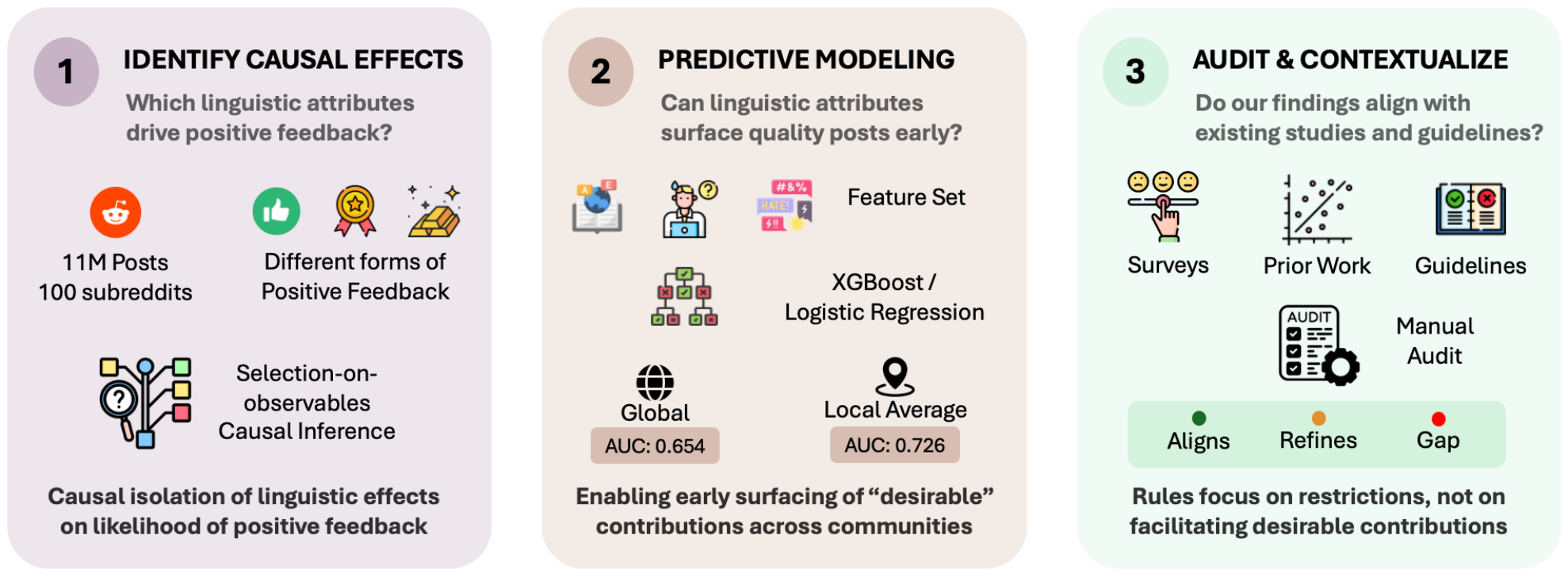}
    \caption{\textbf{Overview of our three-part research approach for studying drivers of positive feedback online.} \textbf{Panel 1 (Identify Causal Effects):} We apply a selection-on-observables causal inference framework to 11M posts from 100 subreddits to isolate the causal impact of linguistic attributes on three forms of positive feedback (upvotes, awards, gold) while controlling for confounding factors like author reputation and timing. \textbf{Panel 2 (Predictive Modeling):} We evaluate whether the linguistic features used in our causal analysis can support real-time detection of high-quality posts through both global and local predictive models, testing their ability to surface high quality contributions before community votes fully accrue. \textbf{Panel 3 (Audit \& Contextualize):} We systematically compare our empirical findings against prior user and moderator surveys, descriptive studies, and community guidelines through a manual audit to identify where our causal estimates align with, refine, or contradict current understanding of what drives positive reception in online communities. Our work combines causal inference, predictive modeling, and comparative analysis to provide both theoretical insights and practical implications for community design and moderation.}
    \Description{Three-panel horizontal layout showing research methodology workflow. Panel 1 displays Reddit logo with various feedback icons including thumbs up, award medal, and gold bar, connected by arrows to a causal inference diagram with branching pathways. Panel 2 shows data processing flow from feature extraction symbols to machine learning model representations, with performance metrics displayed as AUC scores of 0.654 for global and 0.726 for local models. Panel 3 contains survey rating icons, scatter plot visualization, and audit checklist symbols, with color-coded indicators showing alignment in green, refinement in orange, and gaps in red between findings and existing knowledge sources.}
  \end{teaserfigure}

\maketitle

\section{Introduction}

When users make contributions to online communities, subtle linguistic choices---e.g., whether to frame content as a question or statement, use specific versus general language, or adopt formal versus casual tone---can influence its reception. These decisions at an individual level aggregate into community-wide patterns that shape which voices are amplified, whose content gains visibility, and ultimately, how communities evolve~\cite{10.1145/3613904.3642769}. Despite the millions of posts submitted daily across social media platforms, we lack a systematic understanding of the causal mechanisms that drive positive community feedback.

Social media platforms have developed sophisticated reward systems that fundamentally restructure how content is valued and distributed. These systems use collective ratings as proxies for content quality and engagement, with mechanisms that allow users to upvote, like, or award content, and then leverage that feedback to make algorithmic decisions about visibility and prominence~\cite{eckles-senate-hearing}.

\subsection{Opportunities and Challenges in Positive Feedback Mechanisms}

The power of these reward mechanisms extends far beyond simple content ranking. Prior research demonstrates that positive feedback creates reinforcement loops that actively shape future behavior. Being highly upvoted or receiving awards on Reddit encourages users to post content that receives even more positive feedback than they would have otherwise~\cite{10.1145/3706598.3713830}. Similarly, creator recognition features such as ``creator hearts'' on YouTube improve engagement and comment visibility~\cite{10.1145/3706598.3713521}. This pattern reflects broader principles of positive reinforcement validated across offline contexts, from workplace productivity to child development~\cite{gray_psychology_2010,wei_impact_2014,athalye_evidence_2018}. Reward systems do not merely reflect community preferences---they actively incentivize behaviors that align with perceived community values~\cite{weld_making_2021,goyal_uncovering_2024}.

These dynamics create both opportunities and challenges for community governance. Surveys of Reddit moderators reveal strong demand for mechanisms that highlight high-quality content and desirable behavior, rather than focusing solely on removal~\cite{lambert_positive_2024}. Furthermore, research on antisocial behavior shows that undesirable content is much more prevalent than what moderators can address~\cite{park2022measuring}, suggesting a symmetric problem: many promising contributions likely remain underexposed while platforms and moderators focus primarily on content removal~\cite{10.1145/2441776.2441866}. As a result, early-stage detection and lightweight guidance systems could help surface high-quality contributions before organic voting fully occurs, potentially improving platform engagement quality, community health, and newcomer retention by making early participation feel acknowledged.

However, a fundamental gap limits progress towards building such systems: we lack a causal understanding of what drives positive feedback. Current research falls into two categories, each with its own limitations. First, surveys of moderators and users identify what communities say they value~\cite{lambert_positive_2024,weld_what_2022,weld_making_2024}, but surveys may reflect aspirations rather than realized behavioral patterns in practice. Second, descriptive and correlational studies identify attributes in content that are associated with higher engagement~\cite{bar2023analyzing,papakyriakopoulos_upvotes_2023,voinea2024digital,goyal_uncovering_2024}, but correlation conflates language with author reputation, timing, and visibility. We note that this distinction is crucial: knowing that popular posts tend to use specific linguistic pattern is a \emph{reflective} observation, providing limited guidance for actionable \emph{proactive} interventions. Without isolating causal effects, we cannot build reliable detection or guidance tools, or explain why certain interventions succeed or fail.

Addressing this causal identification challenge would enable transformative applications: real-time guidance systems that help users craft more effective contributions, early-detection algorithms that surface quality content before organic and algorithmic patterns solidify, and evidence-based community guidelines that teach effective contribution strategies to users. We believe that Reddit provides an ideal setting for addressing these challenges. Its transparent feedback mechanisms such as upvotes, awards, and gold, massive scale of posts, and community diversity spanning thousands of distinct topics enable a robust causal analysis.

\subsection{Research Questions}
To address these challenges, we pose the following research questions (RQs):

\noindent\textbf{RQ1:} \emph{What linguistic attributes causally drive positive feedback on Reddit?} \begin{enumerate}
    \item[(a)] Which linguistic attributes causally influence the probability that a post receives positive feedback, after conditioning on author reputation and activity as well as timing- and community-level fixed effects? 
    \item[(b)] Are these effects outcome-specific--- i.e., do they differ across feedback channels (free upvotes \emph{v.s.} paid awards and gold)---or are common attributes rewarded across outcomes?
    \item[(c)] How do these effects vary with author status? Do newcomers and established contributors benefit from the same attributes, or do they face different thresholds?
\end{enumerate} 

\noindent \textbf{RQ2:} \emph{Are the curated linguistic attributes also useful for predictive modeling?} \begin{enumerate}
    \item[(a)] To what extent can the attributes studied in RQ1 support real-time discovery of high quality posts---i.e., can models trained on these signals surface promising content that existing voting mechanisms may underrepresent?
    \item[(b)] How does the performance of these models vary across communities?
\end{enumerate}

\noindent \textbf{RQ3:} \emph{How do our findings compare to existing knowledge and community practices?}
\begin{enumerate}
    \item[(a)] How do the empirical, causal drivers of positive feedback identified in our study align with, refine, or contradict claims from prior survey-based and descriptive studies about desirable contributions on Reddit?
    \item[(b)] Do subreddit guidelines currently articulate these empirically supported attributes, and where they do, what gaps or divergences exist between prescriptive guidance and observed drivers of positive feedback?
\end{enumerate}

\subsection{Summary of Contributions}

\subsubsection{Methods:} To answer these research questions, we conduct a large scale empirical study using 11 million posts from 100 communities on Reddit (or \emph{subreddits}) to make three key contributions in this work. 
First, using a quasi-experimental causal inference framework, we identify the attributes from a rich set of linguistic, topical, and semantic features that influence the likelihood of a post receiving positive feedback.
Second, we use predictive modeling on our feature set to model the likelihood of a post being rewarded by the community. Finally, we conduct an audit of existing community guidelines to identify what kind of rules are currently prescribed, and whether they include rules to teach users how to make desirable contributions in context of our findings.

\subsubsection{Findings:} 

Our causal analysis reveals that posts that frame content as broad discussion-generating prompts rather than narrow help requests show $\approx43\%$ higher odds of receiving high scores, while clear, readable writing increases odds by $\approx 40\%$ per standard deviation increase. Conversely, question-heavy posts face $\approx 30\%$ lower odds, toxicity decreases likelihood by $\approx 5\%$, and tentative, hedged language also reduces reception. We identify that specific LIWC features---including causal framing, future focus, and concrete social anchors---positively predict feedback, while excessive informality and dense connective prose are penalized.

These effects vary meaningfully across feedback channels. While discussion-generating framing leads to high score and awards, paid feedback mechanisms (awards and gold) show distinct patterns, with mentions of resources positively associated and power-seeking language penalized for gold. We also find that newcomers face a baseline disadvantage ($\approx6\%$ lower odds of receiving a high score) but benefit disproportionately from clarity and future-focused framing, indicating that linguistic choices can partially compensate for reputation deficits among new community members.

Our predictive models trained on these linguistic features achieve strong performance, outperforming existing prosociality-based approaches by $12\%$. 

Finally, systematic comparison with existing studies reveals that community guidelines currently overwhelmingly focus on enforcement and restrictions rather than teaching the empirically-supported attributes that drive positive feedback, with none of the audited communities providing guidance on the specific linguistic strategies that we find to be causally linked to rewards.

\subsubsection{Implications:} Our work advances HCI research by demonstrating that linguistic choices are measurable, causal drivers of community feedback, and that these linguistic attributes can predict positive reception across communities with high AUC. This would enable platforms to develop real-time guidance tools and early-surfacing mechanisms that identify quality contributions before organic votes accrue. Our findings are also particularly valuable for newcomers, who face baseline disadvantages but benefit disproportionately from clarity and future-focused framing. We also reveal a critical ``policy-practice gap''---while causally-identified reward patterns align with values around quality and prosociality, written community guidelines focus overwhelmingly on restrictions rather than teaching the empirically-supported linguistic and stylistic strategies that drive positive feedback. This suggests platforms and moderators should reconceptualize rules as formation-oriented teaching resources that include exemplars and explicit guidance on effective contribution strategies.

From a theoretical perspective, we extend understanding of online community values by isolating causal linguistic mechanisms from confounding factors like reputation and timing, providing a methodological framework for studying platform behaviors where selection effects typically obscure causality. Our findings call for future field experiments testing whether implementing these linguistic strategies improves long-term user outcomes and community health. Finally, our work opens new avenues for proactive moderation that emphasizes positive reinforcement over purely punitive enforcement.

\section{Related Works}

Understanding what drives positive feedback in online communities sits at the intersection of several research domains.

\subsection{Proactive Community Governance}

As defined by \citet{grimmelmann_virtues_2015}, moderation is ``the set of governance mechanisms that structure participation in a community to facilitate cooperation and prevent abuse.'' While substantial research and tooling exists for reactive moderation approaches focused on \emph{preventing abuse}~\cite{chandrasekharan_crossmod_2019,cai2023hate,wang2025chilling}, significantly less attention has been paid to proactive moderation strategies that \emph{facilitate cooperation} through positive reinforcement. 

This imbalance exists even though both theoretical foundations and empirical evidence supporting the effectiveness of positive reinforcement mechanisms. The psychological foundations for positive reinforcement date back to early behavioral research by \citet{ferster_schedules_1957}, which established that providing positive stimuli following desired behaviors effectively encourages their repetition. This principle has been validated across diverse offline contexts, from workplace productivity \cite{perryer_enhancing_2016} to educational settings \cite{steinberg_impact_1992} and sports performance \cite{wiese_sport_1991}. Importantly, research demonstrates that positive reinforcement can operate both directly on individuals and vicariously through observational learning \cite{bandura_conditions_1971, kazdin_effect_1973}, suggesting that highlighting exemplary behavior can potentially influence broader community norms.

In online contexts, HCI researchers have increasingly recognized the potential of positive reinforcement for improving community dynamics. \citet{kraut_building_2011} identify positive feedback as a key mechanism for encouraging sustained user participation, while \citet{kiesler_regulating_2012} emphasize how highlighting norm-adhering behavior can vicariously reinforce similar patterns among community members. This theoretical foundation has gained empirical support through several platform-specific studies. For example, \citet{wang_highlighting_2021} demonstrated that introducing a simple badge system for New York Times commenters significantly improved both comment quality and frequency, with particularly strong effects for newcomers. \citet{gurjar_effect_2022} showed that when content posted by social media users receives sudden viral attention, the users significantly increase posting frequency and alter their content to more closely match the viral post. \citet{10.1145/3706598.3713830} showed that positive feedback mechanisms on Reddit causally improve users' future content quality and reduce removal rates, while \citet{10.1145/3706598.3713521} showed that using ``creator hearts'' as a positive reinforcement strategy for desirable comments on YouTube improves engagement and comment visibility. 

Despite this evidence, moderators report limited access to tools supporting positive reinforcement strategies. The survey of Reddit moderators by \citet{lambert_positive_2024} found an explicit demand for better features enabling moderation through positive reinforcement rather than purely punitive approaches. Similarly, \citet{liu2025needling} discovered that moderators using their toxicity visualization tool expressed interest in incorporating positive reinforcement techniques into their workflows.

\emph{Our work addresses these gaps by identifying attributes that drive positive feedback on Reddit, enabling the development of evidence-based positive reinforcement tools that can both detect promising content early and guide users toward effective contribution strategies.}

\subsection{Content Quality Assessment and Modeling Community Feedback Mechanisms}

Research on content quality and community feedback has primarily focused on understanding what characteristics correlate with or describe positive reception. 

The survey by \citet{weld_making_2024} provides a comprehensive taxonomy of what Reddit users claim to value in the communities they participate in. Their work identifies quality content and engagement as key themes that users value. In parallel, the survey by \citet{lambert_positive_2024} reveals that Reddit moderators similarly value engaging, prosocial, and high quality content.

On the other hand, many studies focus on empirical approaches to predicting content success, operationalize certain measures, or explain community dynamics. For example, researchers have developed models using temporal and environmental patterns~\cite{szabo2010predicting,shimgekar2025detecting} and engagement cascades~\cite{cheng2014can}. \citet{mitra_language_2014} studied the factors which led to successfully funding a crowdfunding projects on Kickstarter, but do not offer causal claims. \citet{lambert_conversational_2022} studied what attributes in content enable conversations on Reddit to be resilient to adverse events such as content removals. Similarly, the framework and models introduced by ~\citet{bao_conversations_2021} have been used widely for measuring desirable contributions, although it captures only a subset of factors that drive positive feedback, i.e., prosociality. The work of \citet{10.1145/3613904.3642769} focuses on language at an averaged sentence embedding level to demonstrate that online communities face a trade-off between growth and linguistic distinctiveness. This further suggests that community evolution fundamentally reshapes the linguistic landscape that can determine what content receives positive reception. These approaches primarily focus on prediction and measuring correlation rather than understanding causal mechanisms. 

Prior research that has focused on answering causal questions have studied downstream effects of receiving positive/negative feedback~\cite{chandrasekharan_quarantined_2022,papakyriakopoulos_upvotes_2023} or high engagement~\cite{gurjar_effect_2022}, which means an in-depth understanding of what actually causes this feedback is missing. Similarly, research on mental health outcomes has used quasi-experimental designs to examine the effectiveness of the Papageno effect~\cite{domaradzki2021werther} for suicidal ideation on Twitter (now X)~\cite{yuan_mental_2023}, and the effect of news exposure on well-being outcomes~\cite{pal2026hidden}. Different from our work, these studies focus on behavioral outcomes rather than linguistic mechanisms. 
More similar to our setting, \citet{zhang_conversations_2018} applied causal inference to study conversational failure, i.e., isolate the effects of early warning signals on conversations that end in undesirable outcomes.

\emph{While extensive research documents correlational relationships between content features and positive reception or other downstream effects, a rigorous causal analysis of drivers remains absent, representing a fundamental gap that our causal inference approach addresses.}

\subsection{Proactive Intervention Systems and User Guidance Tools}

Recent work has demonstrated both the feasibility and effectiveness of proactive intervention systems for online communities. \citet{Habib_Musa_Zaffar_Nithyanand_2022} showed that the evolution of Reddit communities toward hateful or dangerous behavior can be predicted months in advance, providing administrators with a scientific rationale for proactive moderation decisions and demonstrating that such approaches are indeed feasible. Building on the foundation of proactive approaches, \citet{schluger_proactive_2022} investigated existing proactive moderation practices on Wikipedia Talk Pages, finding that moderators already engage in proactive behaviors---such as preemptively intervening in conversations to keep them on track---but lack adequate technical support for these efforts. Similarly, \citet{doi:10.1073/pnas.1813486116} showed that posting a short ``sticky'' comment with community rules in \emph{r/science} discussions increased rule compliance in first-time commenters by $\approx8.4$ percentage points and also boosted newcomer participation by $\approx70\%$, showing that making norms visible both curbs unruly behavior and draws in more newcomers. 

Several systems have emerged to support positive community outcomes through different intervention strategies. \citet{park_supporting_2016} developed CommentIQ to help news outlet moderators identify and highlight high-quality comments, recognizing that promoting good contributions can improve overall community quality beyond simply removing problematic content. Similarly, \citet{masrani2023slowing} explored temporal interventions, finding that adding friction through message-sending delays led to more thoughtful messaging, though participants experienced both frustration with the constraint and appreciation for its mindfulness-promoting effects.

Most directly relevant to our work, \citet{horta2025post} developed Post Guidance, a system that proactively guides users during post composition on Reddit through community-specific rules that can surface messages, prevent submissions, or flag content for review. 

\emph{Our work extends this line of research by identifying empirically-grounded linguistic features that could enhance such proactive systems with specific, causal drivers of positive community reception. We also use our feature set to predict the likelihood of a post receiving positive feedback.}

\section{Data Curation, Methods, and Study Design}

\begin{figure*}
    \centering
    \includegraphics[width=\linewidth]{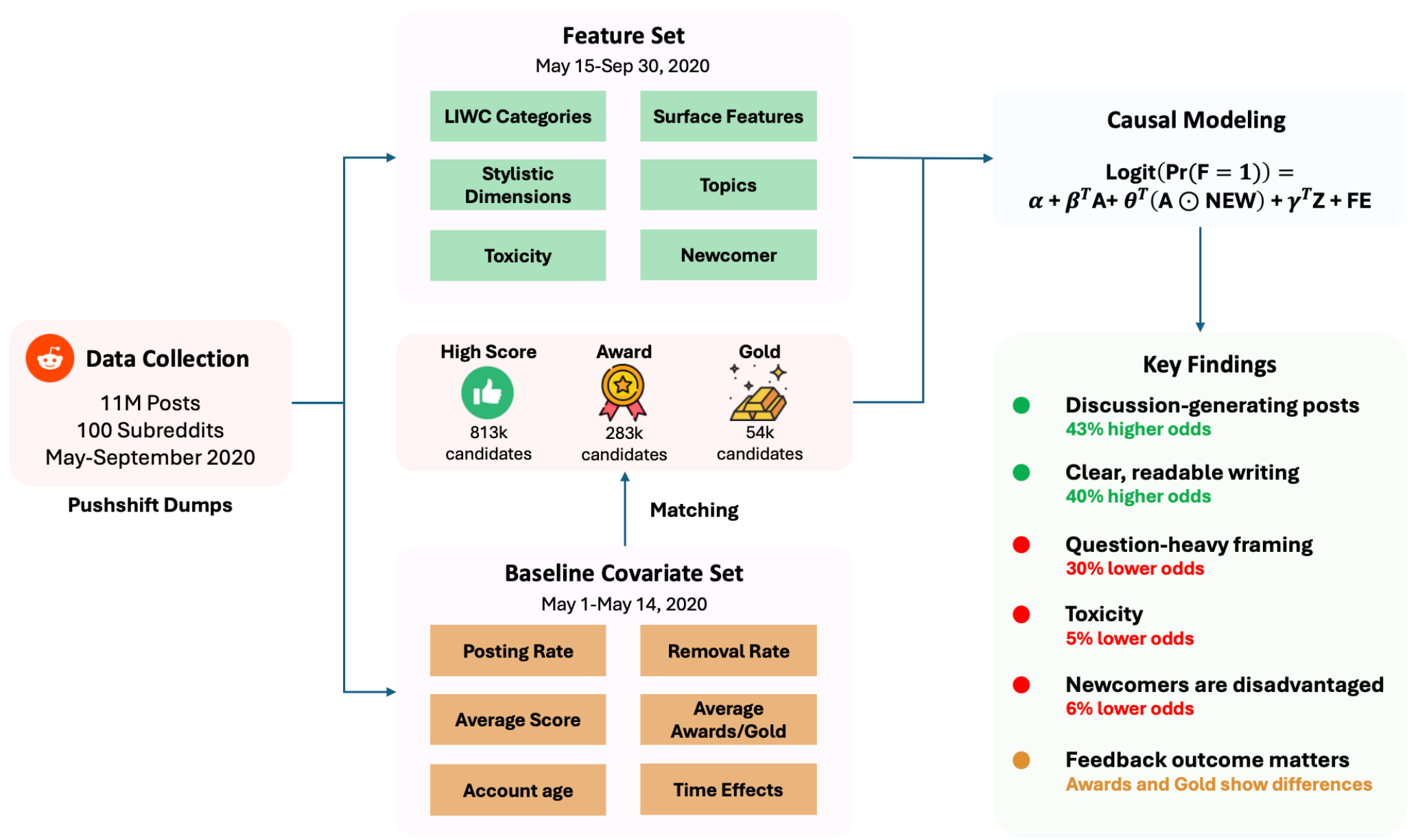}
    \caption{\textbf{Overview of causal inference framework for identifying linguistic drivers of positive feedback on Reddit.} We analyze 11M posts from 100 subreddits (May-September 2020) using a selection-on-observables approach that isolates causal effects of linguistic attributes from confounding factors. The framework combines rich linguistic feature extraction (100+ attributes spanning LIWC categories, surface markers, semantic style dimensions, topics, and toxicity) with baseline covariates capturing author reputation and activity patterns from a 14-day pre-period. Matching based on subreddits-wise risk stratification creates balanced comparison groups, while fixed effects control for community norms and temporal variation. The logistic regression model estimates causal effects on three positive feedback outcomes: high scores (top $25\%$ within subreddit-month, $813\text{k}$ candidates), awards ($283\text{k}$ candidates), and gold ($54\text{k}$ candidates). Key causal findings include: discussion-generating posts have $43\%$ higher odds of high scores compared to narrow help requests; clear, readable writing increases odds by $40\%$; question-heavy framing reduces odds by $30\%$; and toxicity consistently decreases reception. Newcomers face baseline disadvantages by $6\%$ but benefit disproportionately from clarity and future-focused language.}
    \Description{Horizontal workflow diagram showing data processing pipeline from left to right. Left panel shows Reddit logo with data collection statistics. Center contains two feature grids: upper grid has six green boxes representing linguistic feature categories, lower grid has six orange boxes representing baseline covariates from May 1 to May 14, 2020. Arrows flow from both grids toward right side showing causal modeling process with risk stratification step. Right side displays logistic regression equation and three outcome types with icons: thumbs up for high score with 813k candidates, medal for award with 283k candidates, and gold bar for gold with 54k candidates. Far right panel lists key findings with colored bullet points indicating positive effects in green, negative effects in red, and newcomer-specific effects in orange.}
    \label{fig:causal-pipeline}
\end{figure*}

Our goal is to quantify how observable post-level attributes---e.g., when the submission is made, who made it (i.e., the author's baseline activity, tenure in the community, and prior karma), and its content---shape the likelihood of receiving positive feedback. Formally, we ask: By how much does the probability of earning a high score, an award, or a gold change when a given attribute increases or decreases, holding all other observed factors constant? Because a single post cannot exist simultaneously with and without that attribute, the counterfactual outcome is unobserved. We therefore adopt a \emph{selection-on-observables framework}~\cite{cochran1968effectiveness,EGGERS_HAINMUELLER_2009,sekhon2009opiates,stuart2010matching} for causal inference. In our setting, this means that, conditional on observed author history/activity and on community/timing fixed effects (i.e., the observed covariates), we assume that there are no unmeasured factors that jointly affect the focal linguistic attribute of interest (i.e., the treatment) and the likelihood of positive feedback (i.e., the outcome)---so the remaining variation in that attribute is as good as random with respect to the outcomes.

\edit{We employ selection-on-observables because stronger causal designs are infeasible for our research. Randomized experiments (RCT) would require manipulating user content (unethical and/or ecologically invalid), while difference-in-differences (DiD) and regression discontinuity (RDiT) require specific treatment structures---discrete timing, exogenous instruments, or sharp discontinuities---that don't exist for continuously varying linguistic choices like in our setting. Our approach requires assuming \emph{conditional independence}, where after controlling for author history, timing, and community context, residual linguistic variation is as-good-as-random. While this assumption is untestable, our design represents a substantial methodological advancement over descriptive studies by isolating linguistic effects from major confounds, providing \emph{conditional causal estimates} rather than correlational associations. We therefore highlight that our estimates should be interpreted as \emph{conditional causal effects under stated assumptions rather than assumption-free causal claims}.} See Section 7.6.3 for further discussion.

Our data comes from the Pushshift archives of Reddit data~\cite{baumgartner_pushshift_2020} from May 1,
2020 to Sep 30, 2020, mirroring the study window of \citet{10.1145/3706598.3713830}. We chose this window to maintain consistency with prior work on studying positive reinforcement on Reddit, and to ensure that all commenting and voting activity in these subreddits has stabilized by the time of this study. This ensures that the community feedback we model is representative. Since our research did not involve any user participation, all data was publicly available on Reddit, and we ensured to maintain anonymity in our data to ensure user privacy, our work did not require approval from an institutional review board (IRB). We describe our feedback outcomes and candidate pool construction in the sections below.

\subsection{Positive Feedback Outcomes}

Our analysis uses three indicators of approval that were active on Reddit during the observation window: \emph{score}, \emph{awards}, and \emph{gold}. We center our analyses on \emph{score} (roughly net \# of upvotes - \# of downvotes), a free-for-all voting mechanism available in every subreddit, making it a broad measure of community approval.

To complement this broad signal, we also consider two paid forms of positive feedback: \emph{awards} and \emph{gold}. These signals are comparatively sparse in practice but carry a higher bar of endorsement because they incur a monetary cost. Reddit Gold was a popular yet infrequently used mechanism~\cite{wang_highlighting_2021}, and was removed in 2023~\cite{weatherbed_2024_reddit}. Awards were also removed in 2023\footnote{\href{https://reddit.com/r/reddit/comments/14ytp7s/reworking_awarding_changes_to_awards_coins_and/}{https://reddit.com/r/reddit/comments/14ytp7s/reworking\_awarding\_changes\_to\_\\awards\_coins\_and/}} and later reinstated.\footnote{\href{https://reddit.com/r/reddit/comments/1css0ws/we_heard_you_awards_are_back/}{https://reddit.com/r/reddit/comments/1css0ws/we\_heard\_you\_awards\_are\_back/}} However during our observation window, both mechanisms were available and we therefore include them in our analysis.

Operationally, we consider a post ``highly upvoted'' if its \emph{score} falls in the top 25th percentile of post-level scores within a subreddit-month pair, ``awarded'' if it received at least one award, and ``gilded'' if it received at least one gold.

Note that, consistent with prior work~\cite{10.1145/3706598.3713830,goyal_uncovering_2024}, we interpret \emph{score}, \emph{awards}, and \emph{gold} as a reaction signal of approval by the community rather than a guarantee of accuracy or quality. There is a possibility that these mechanisms do not necessarily represent a high-quality contribution, and that there are more important signals of ``desirability'' than getting highly upvoted (see Section 7.5).

\begin{figure*}
    \centering
    \includegraphics[width=\linewidth]{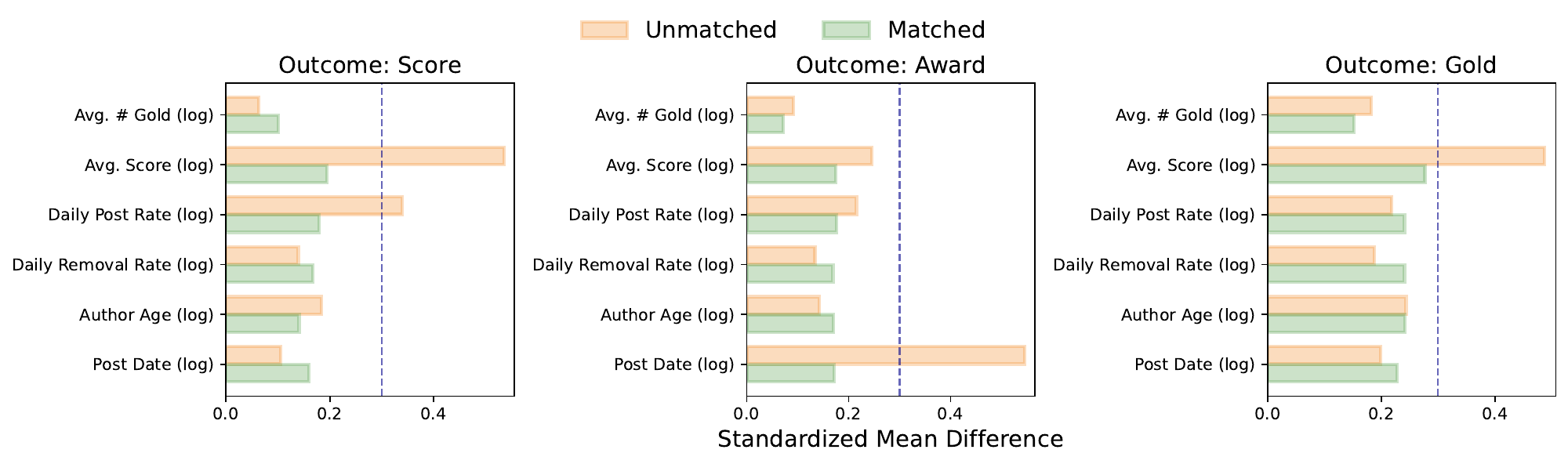}
    \caption{Standardized mean differences (SMD) for covariates in $\mathbf{Z}$, averaged across retained strata. Risk stratification generally reduces imbalance; where pre-stratification imbalance was already low, changes are small. For all outcomes, the mean post-stratification SMD is below the 0.30 threshold, indicating adequate balance.}
    \Description{Three-panel horizontal bar chart comparing covariate balance before and after risk stratification matching. Each panel represents one outcome type with six covariates displayed as paired horizontal bars extending from zero to approximately 0.4 on the standardized mean difference scale. Orange bars show unmatched differences, green bars show matched differences. Vertical dashed line marks the 0.30 adequate balance threshold. Most green bars fall below this threshold while several orange bars exceed it, demonstrating that the matching procedure successfully reduces covariate imbalance across all three outcome types. The pattern is consistent across Score, Award, and Gold outcomes, with the largest improvements typically seen for Average Score and Daily Post Rate covariates.}
    \label{fig:smd_mean}
\end{figure*}

\subsection{Candidate-pool Construction}
 
Since \emph{gold} is the rarest of our feedback signals, we construct our study data using the 100 subreddits that awarded the most gold during the observation window, ensuring sufficient positive cases for all three outcomes.

Following \citet{10.1145/3706598.3713830}, we partition our overall observation window into a \emph{baseline window} and a \emph{sampling window}. The overall observation window is May 1--Sep 30, 2020. The 14-day baseline spans May 1--May 14 and is used to compute pre-feedback author covariates (e.g., posting and removal rates, typical score/awards, account age). \edit{A 14-day baseline window is standard in prior social-media based causal inference works~\cite{yuan_mental_2023,10.1145/3706598.3713830}.} The sampling window spans May 15--Sep 30, 2020 during which candidates are constructed and outcomes are evaluated.

For each outcome and subreddit, we form two sets within the sampling window:
\begin{itemize}
    \item \textbf{Positive-feedback candidates:} posts that satisfy the outcome definition described above (``highly upvoted,'' ``gilded,'' or ``awarded'').
    \item \textbf{Control candidates:} posts from the same subreddit and calendar day that do not satisfy the outcome definition. For \emph{score}, we draw controls from the bottom two quartiles and omit the third quartile to avoid near-threshold posts that could blur contrasts and interpretations. Same-subreddit, same-day controls mitigate diurnal and community-level activity swings.
\end{itemize}

To guarantee baseline information, we require that each author posted at least once during the baseline window. Control samples are drawn at a $3{:}1$ ratio relative to positive-feedback samples. This procedure yields approximately $813\text{k}$ candidates for \emph{score}, $283\text{k}$ for \emph{awards}, and $54\text{k}$ for \emph{gold}. We next describe how we create comparable risk strata to reduce confounding before modeling.

\subsection{Baseline Covariates ($\mathbf{Z}$)}

In our selection-on-observables framework, we want to control for attributes that may affect our attributes of interest (i.e., the linguistic attributes we use as treatments like toxicity) and the outcome variable. We therefore choose the following non-linguistic covariates \(\mathbf{Z}\) (log-scaled where appropriate):

\begin{itemize}
    \item Daily posting rate in the baseline window.
    \item Daily removal rate in the baseline window.
    \item Average score on the author's posts in the baseline window.
    \item Average number of awards/golds on the author's posts in the baseline window.
    \item Author's Reddit account age at the time of posting.
    \item Secular time trend (days since May 1, 2020).
\end{itemize}

These capture prior activity, reputation, tenure, moderation exposure, and time trend measured in the 14-day baseline window. Having defined the non-linguistic baselines $\mathbf Z$, we next introduce the pre-feedback linguistic feature families $\mathbf A$ that are the focus of our analysis. While it is possible that there are other covariates, such as demographic biases or factors such as real-world ties, we cannot observe them directly from Reddit data and we therefore do not consider them in our analysis.

\subsection{Core Feature Set}

We wish to model the probability that a contribution receives strong positive feedback as a function of a rich set of linguistic attributes that capture lexical content, surface form, latent semantics, and psychosocial textual properties. Below we define the feature sets we compute for our modeling objective.

\subsubsection{Linguistic Feature Set ($\mathbf A$):}
For each post, we compute a vector of textual attributes $\mathbf{A}$ spanning surface cues, psycholinguistic properties, latent topics, semantic style, and toxicity. All features are computed directly from the text.

\begin{itemize}
    \item \textbf{LIWC-2015:} Using Linguistic Inquiry and Word Count~\cite{pennebaker2001linguistic}, we compute the percentage of tokens matching each psycholinguistic category (e.g., affect, social processes, cognitive processes, function words), where values lie in \([0,100]\). LIWC offers interpretable, theory-grounded correlates widely used in psychology and HCI~\cite{10.1145/2858036.2858535,lester2010content,10.1145/3501247.3531572,10.1145/3544548.3581318}.
    \item \textbf{Surface attributes:} We compute sentiment via the VADER compound score \([\!-1,1]\)~\cite{hutto2014vader}, Flesch reading ease~\cite{kincaid1975derivation}, interrogative style (fraction of sentences in the post ending in ``?''), and politeness using ConvoKit~\cite{chang-etal-2020-convokit}. These cues summarize clarity, tone, and rhetorical posture that authors can adjust at posting time.
    \item \textbf{Latent topics:} We estimate topic probabilities with Latent Dirichlet Allocation (LDA)~\cite{blei2003latent} using \edit{\(K \in \{5, 10, 15, 20, 25\}\) topics}; topic probabilities lie in \([0,1]\) and sum to 1 per post, capturing coarse content structure. LDA has been used extensively in prior social media research as it produces stable and interpretable topics~\cite{chancellor2016quantifying,ernala2017linguistic,10.1145/3613904.3642078}.
    \item \textbf{Semantic style:} We embed each post with Sentence-BERT (\texttt{all-MiniLM-L6-v2})~\cite{reimers_sentence-bert_2019} into a 384-dimensional vector and project onto the top 10 principal components (PCA)~\cite{abdi2010principal}. These orthogonal components summarize high-level discourse/phrasing directions. We measured percentage of explained variance by computing between 5 and 20 principal components from our data, and found a saturation at 10 components at roughly 56\%. 
    \item \textbf{Toxicity:} We score toxicity using Detoxify~\cite{Detoxify}, a popular toxicity detection model, returning a score in \([0,1]\).
\end{itemize}

Our goal is to estimate the \emph{marginal effect of observable textual attributes} while holding constant author baselines and community/time conditions introduced above.

\subsubsection{Indicator for newcomers (\(\textsc{NEW}\)):}
We include a binary indicator \(\textsc{NEW}\in\{0,1\}\) for newcomers (account age \(< 90\) days). While \(\textsc{NEW}\) is not a text feature, we use it to form interactions with \(\mathbf{A}\) in the estimation model to test whether linguistic effects differ for newcomers versus established contributors.

\subsection{Feature Engineering and Preprocessing}

After collecting the baseline covariates ($\mathbf{Z}$), the linguistic feature set ($\mathbf{A}$) and the newcomer indicator (\textsc{NEW}), we perform some feature engineering and data preprocessing steps which we describe below.

\subsubsection{LIWC Residual Umbrellas:} LIWC categories are hierarchical, where each coarse ``umbrella'' constructs such as ``\emph{cogproc},'' ``\emph{affect},'' etc. contain finer-grained children, for e.g., \emph{insight}, \emph{cause} for ``\emph{cogproc}'' and \emph{posemo}, \emph{negemo} for ``\emph{affect}.'' \edit{Our initial experimentation revealed that when umbrella and children terms are used together as predictors, the umbrella term is often highly correlated with the sum of its children, which leads to multicollinearity and ambiguous coefficients (e.g., unstable magnitudes or sign flips as umbrella and its child attributes compete to explain the same variance). Therefore, to maintain interpretability and robustness in our methodology, we want the model to include an umbrella-level signal that provides signal \emph{beyond} what the children already capture.} 

To do so, for each umbrella $U$ with children $C_U$ we regress $U$ on its children across posts and replace the umbrella with the residual. Concretely, for each post \(i=1,\dots,n\) we fit a linear model of the umbrella on its children:
\begin{align}
    U_i \;=\; \alpha_U \;+\; \sum_{c\in C_U} w_{Uc}\, C_{ic} \;+\; \varepsilon_{Ui},
\end{align}
where \(U_i\) is the umbrella score for post \(i\) and \(C_{ic}\) is the score of child \(c\) for that post, and $\alpha_U$ is the intercept term. The \emph{residual umbrella} is then defined as the regression residual:
\begin{align}
    \emph{other\_}U_i \;\equiv\; \varepsilon_{Ui}
\;=\; U_i \;-\; \big(\hat{\alpha}_U + \sum_{c\in C_U} \hat{w}_{Uc}\, C_{ic}\big).
\end{align}

Intuitively, \(\emph{other\_}U_i\) measures how much of the umbrella \(U\) remains after its named subcategories are accounted for. In practice, this orthogonalizes umbrellas to their children, which helps to reduces collinearity and yields more stable and interpretable effects.

\subsubsection{Topics as Composition:} As mentioned in the previous section, we use LDA for topic modeling, However, LDA returns per-post topic proportions \(\{p_{ik}\}_{k=1}^K\) that are nonnegative and sum to 1. Treating these raw proportions as ordinary predictors can create problems as increasing one topic necessarily decreases others due to the unit-sum constraint, which again induces spurious negative correlations and collinearity. To analyze topic mix in a way that accounts for its compositional nature, we transform the proportions with the centered log-ratio (CLR), which expresses each topic relative to the \emph{geometric mean} of the others.

Concretely, for post $i$ with $K$ topics, we add a small constant $\varepsilon$ (to handle zeros), compute the row-wise geometric mean $g_i$, and then the CLR coordinates:
\begin{align}
    g_i \;=\;\Big(\prod_{k=1}^{K} (p_{ik}+\varepsilon)\Big)^{1/K}, \qquad
\mathrm{clr}_{ik} \;=\; \log\!\frac{p_{ik}+\varepsilon}{g_i}.
\end{align}
Since CLR components sum to zero, to avoid a singular design sub-matrix we drop one CLR column before estimation. Coefficients on \(\mathrm{clr}_{ik}\) will then be interpreted as \emph{relative} effects, i.e., a one–standard-deviation increase corresponds to moving distribution \emph{towards} topic \(k\) and \emph{away from} the geometric mean of the other topics.

\subsubsection{Standardization:}
Finally, we \(z\)–score all numeric features in both the linguistic  \(\mathbf A\) and baseline covariate sets \(\mathbf Z\) so that all effect sizes we report correspond to a \textit{one–standard–deviation} (``1-SD'' hereafter) change in that feature.

\subsection{Risk Stratification and Balance}

Within each subreddit $s$ and for each outcome $T$, we fit a \emph{non-linguistic} model on $\mathbf{Z}$ to estimate a \emph{baseline outcome risk score}
\begin{align}
r(\mathbf{Z}, s) \;=\; P\!\left(T{=}1 \mid \mathbf{Z}, s\right),
\end{align}
using an AdaBoost classifier~\cite{freund1999short}. \edit{Prior causal inference work has used simpler models such as the averaged perceptron learning algorithm~\cite{de2016discovering} and logistic regression~\cite{yuan_mental_2023} for risk stratification and propensity score matching, however recent works such as \citet{10.1145/3706598.3713830} use Adaboost as it flexibly capture nonlinearities and interactions in the covariates, and we therefore use Adaboost in our work.} We intentionally exclude linguistic features \(\mathbf{A}\) from the risk model so that linguistic variation remains to explain residual differences \emph{within} well-balanced strata. Using a treatment-style propensity score similar to a potential-outcomes framework~\cite{rubin_causal_2005} would require a single well-defined treatment. We therefore do not use propensity score as our linguistic attributes are not binary ``treatments'' in the sense of a potential outcomes framework.

We bin our outcome risk scores $r(\mathbf{Z}, s)$ into deciles and retain strata with adequate overlap (minimum ten positive-feedback and control candidates). To assess covariate balance between positive and control candidates \emph{within each stratum}, we compute standardized mean differences (SMD) for all covariates in $\mathbf{Z}$ and, following prior work, discard strata whose \emph{mean} SMD across \(\mathbf{Z}\) exceeds \(0.30\)~\cite{kampenes_systematic_2007,kiciman_using_2018,10.1145/3706598.3713830} (see \autoref{fig:smd_mean}). This design yields many-to-many, within-subreddit comparisons among posts with similar baseline risk of receiving positive feedback.

\paragraph{Why subreddit-wise stratification?}
Communities differ in size, activity, norms, and generosity, so baseline \emph{outcome risk} is not proportionate across subreddits. For example, a post with risk $r=0.30$ in \emph{r/science} is not directly comparable to a post with $r=0.30$ in \emph{r/Minecraft}, because $r(\mathbf Z,s)$ is learned against subreddit-specific distributions. A single global model that forms strata across all communities would therefore pool incomparable posts and undermine balance. By stratifying \emph{within} subreddits, we compare positive-feedback and control candidates that are similar on $\mathbf Z$ under the same local risk regime, and therefore the subsequent fixed effects would then identify linguistic effects from within-cell variation.

\subsection{Causal Estimation Model}

For each outcome \(\mathcal{F} \in \{\emph{score}, \emph{award}, \emph{gold}\}\), we estimate a generalized linear model with a \emph{logit} link function:
\begin{equation}
\begin{aligned}
\operatorname{logit}\,&P(\mathcal{F}_i{=}1) \;=\;
\;\alpha \;+\; \beta^\top \mathbf A_i \;+\; \theta^\top(\mathbf A_i \odot \textsc{NEW}_i) \;+\; \gamma^\top \mathbf Z_i \\
\;+\; &\underbrace{\mu_{\,s(i),\,d(i)}}_{\text{subreddit}\ \times\ \text{risk-decile FE}}
\;+\; \underbrace{\delta_{\,\mathrm{calday}(i)}}_{\text{calendar-day FE}}
\;+\; \underbrace{\eta_{\,\mathrm{hour}(i)}}_{\text{hour-of-day FE}}\,.
\end{aligned}
\label{eq:logit_model}
\end{equation}

Here, \(i\) indexes posts; \(s(i)\) is the subreddit and \(d(i)\) its risk decile. The \(\textsc{NEW}_i\in\{0,1\}\) indicator marks newcomers (account age \(<90\) days), and \(\mathbf A_i\odot\textsc{NEW}_i\) captures heterogeneity in language effects between newcomers and veterans, i.e., whether newcomers have any unique additive effects because by virtue of them being new to the platform.

We note interpretations from our modeling equation \eqref{eq:logit_model} below:
\begin{itemize}
    \item $Pr(\mathcal{F} = 1)$ is the conditional probability, given our linguistic feature set ($\mathbf{A}$), baseline covariates ($\mathbf{Z}$), our fixed effects ($\textup{FE}$), that a post receives the chosen positive-feedback signal $\mathcal{F}$ (e.g., whether the post received at least one award).
    \item $\beta_k$ (main effect coefficients) is the \emph{conditional log‑odds change} associated with a one‑unit increase in feature $A_k$, holding all other variables fixed. The \emph{odds ratio} is $\textup{OR} = \exp(\beta_k)$.
    \item $\theta_k$ (interaction term coefficients) modifies the effect of features in $\mathbf{A}$ for newcomers. Therefore, for feature $A_k$ if veterans have effect $\beta_k$, then newcomers have an effect $\beta_k + \theta_k$.
    \item The fixed-effects ($\textup{FE}$) block absorbs systematic community and temporal variation including baseline outcome risk within communities, calendar-day FE for day-level shocks, and hour-of-day FE for diurnal patterns.
\end{itemize} 

We compute one-way cluster-robust standard errors at the \emph{subreddit} level to allow arbitrary within-subreddit correlation (shared norms, moderation shocks). Furthermore, we control for false discovery rate (FDR) of features using Benjamini–Hochberg~\cite{benjamini1995controlling} correction applied separately to main language effects \((\beta)\) and interaction effects \((\theta)\).  We report $q-$values arising after correction of the original $p-$values, and treat $q < 0.05$ as the threshold for ``significant discoveries.'' We choose BH–FDR over Bonferroni correction as the number of features in our model is large, and Bonferroni would be overly conservative, which could mask real effects. Moreover, given the exploratory nature of our RQs, controlling the proportion of false discoveries is more appropriate than forcing zero false positives. Finally, since features are \(z\)–scored, we report odds ratios (OR) for a 1–SD increase. 

Now that we have outlined our rich set of features and our causal modeling objective, we move to answering our research questions.

\section{RQ1: Studying Causal Effects of Linguistic Attributes on Positive Feedback}

In this section, we present results for RQ1 which investigates the linguistic, topical, and behavioral attributes that influence the probability of Reddit posts receiving strong positive feedback. As mentioned previously, the analysis below is based on the significant results ($q < 0.05$ after Benjamini-Hochberg FDR correction) from logistic regression models fitted on the matched samples.

\subsection{RQ1(a): Which attributes raise or lower the likelihood of positive feedback?}

Our analysis reveals many linguistic and semantic features that statistically significantly influence the probability of receiving positive feedback on Reddit. We organize these findings by feature type and discuss their implications. \textbf{We focus on the outcome of ``high score'' in RQ1(a), and postpone discussion of awards and gold to \emph{RQ1(b)}.}

\begin{table*}[t]
\sffamily
\small
\centering
\caption{Semantic dimensions that predict top–quartile score across the 100 subreddits. Odds ratios ($=\exp(\beta_k)$) are estimated from a logistic model fit for the ``high-score'' outcome. Odds ratios above 1 indicate dimensions that increase the likelihood of community reward, while ratios below 1 indicate decreased likelihood. Arrows (↑↑ = increases odds, ↓↓ = decreases odds) summarize practical direction of implications for communities. Significance levels for $\beta_k$: \textbf{***}\,$q<0.001$, \textbf{**}\,$q<0.01$, \textbf{*}\,$q<0.05$.}\label{tab:pc-dimensions}
\Description{Ten-row table showing principal component analysis results with columns for component number, odds ratio, statistical significance, interpretation, and practical implications. Odds ratios range from 0.93 to 1.43, with PC1 showing the strongest positive effect at 1.43 for broad engagement prompts versus narrow help requests. Three components show statistically significant effects marked with triple asterisks: PC1 at 1.43, PC6 at 0.94 for casual social updates, and PC8 at 0.93 for time-stamped content. PC10 shows single asterisk significance at 0.95 for serendipitous shares. Color coding distinguishes positive effects in green, negative effects in red, and neutral effects in gray. Most components show negative or neutral associations with high scores, with only PC1 and PC7 showing positive associations above 1.0. The rightmost column uses directional arrows to indicate whether each dimension increases or decreases odds of receiving high scores.}

\begin{tabular}{lcccl}
\rowcolor{gray!15}
\textbf{PC} &
\textbf{Odds Ratio} & 
\textbf{$q$-sig} &
\textbf{Interpretation from Exemplars} & 
\textbf{Implication for High Score Reward} \\
\cmidrule(lr){1-1} \cmidrule(lr){2-2} \cmidrule(lr){3-3} \cmidrule(lr){4-4} \cmidrule(lr){5-5}
PC1  & \cellcolor{ForestGreen!15}1.43 & *** & Narrow help–requests \emph{vs.} broad engagement prompts & \cellcolor{ForestGreen!15}↑↑ Broad, discussion-generating posts \\
PC2  & \cellcolor{Crimson!15}0.99 & -  & Text-only discussion \emph{vs.} visual/creative showcases & \cellcolor{Crimson!15}↓↓ Text-only content less favored \\
PC3  & \cellcolor{gray!15}1.00 & - & Personal‐struggle narratives \emph{vs.} creative achievements & \cellcolor{gray!15} --- \\
PC4  & \cellcolor{Crimson!15}0.98 & - & Dense, list-like technical exposition & \cellcolor{Crimson!15}↓↓ Complex enumerations \\
PC5  & \cellcolor{Crimson!15}0.98 & -  & Off-topic digressions / meta commentary & \cellcolor{Crimson!15}↓↓ Meta or digressive content \\
PC6  & \cellcolor{Crimson!15}0.94 & *** & Casual social updates \emph{vs.} substantive discussion & \cellcolor{Crimson!15}↓↓ Brief ``status'' posts \\
PC7  & \cellcolor{ForestGreen!15}1.03 & - & Enthusiastic, community-oriented calls & \cellcolor{ForestGreen!15}↑↑ Collective enthusiasm \\
PC8  & \cellcolor{Crimson!15}0.93 & *** & Event-dated / time-stamped content & \cellcolor{Crimson!15}↓↓ Heavily time-bound posts \\
PC9  & \cellcolor{Crimson!15}0.97 & - & Literal descriptions \emph{vs.} reflective commentary & \cellcolor{Crimson!15}↓↓ Purely literal recounting \\
PC10 & \cellcolor{Crimson!15}0.95 & * & Serendipitous “found this” shares & \cellcolor{Crimson!15}↓↓ Chance discoveries over curated insight \\
\bottomrule
\end{tabular}
\end{table*}

\subsubsection{Semantic Style Dimensions (PCs)}

Many significant effects emerge from the \emph{semantic style dimensions} captured by our principal components analysis (PCA) of sentence embeddings. These Principal Components (PCs) represent orthogonal directions in the high-dimensional semantic space that capture distinct discourse styles and content types.

To identify what each of these dimensions encodes, we take the dot product between the eigenvector of each PC direction $\mathbf{v}_k$ and the SentenceBERT embedding vector $\mathbf{x}_i$ for each post $i$, to obtain the PC score: 

\begin{equation}
    \text{PC}_{i,k} = \mathbf{x}_i^T \mathbf{v}_k\,\,\,\,\forall \,\,k \in \{1, \cdots, 10\}.
\end{equation}

This score $\text{PC}_{i,k}$ is a scalar value representing how much that post ``aligns with'' or ``loads on'' that particular semantic dimension. Strong positive scores indicate greater alignment with this dimension, while strong negative scores indicate weak or even opposite alignment with the dimension.

We then pick the top-30 and bottom-30 posts by PC scores for each principal component and perform manual inspection as well as cross-validation with GPT-5 model~\cite{openaiIntroducingGPT5} (\texttt{gpt-5-2025-08-07}) to interpret what the most aligned and unaligned posts within each dimension indicate. The prompt used can be found in \autoref{app:llm-prompt}. \edit{We found that GPT-5 interpretations matched manual inspection, strengthening our confidence in the dimensions extracted.} By selecting the posts with the most extreme PC scores along each semantic axis, we hope to capture true signal in terms of what kind of content aligns with these dimensions.

\textbf{Findings:} \autoref{tab:pc-dimensions} shows that a subset of semantic style dimensions exert clear, statistically reliable effects on whether a post reaches top–quartile score, while others provide \emph{directional}, but not decisive signals. Since features are $z$–scored, odds ratios reflect the change in odds for a 1–SD shift along each semantic axis, holding author baselines and community/time conditions constant.

The strongest positive effect is \emph{PC1} ($OR=1.43, q < 0.001$), which indicates that posts framed as broad, discussion-generating prompts as opposed to narrow help requests are associated with $\approx43\%$ higher odds of community reward. This highlights the power of framing---inviting dialogue and shared sense-making provides a greater pay off to posters. On the other hand, three dimensions are reliably disfavored. \emph{PC6} ($OR = 0.94, q < 0.001$) indicates that casual, status-update style posts underperform relative to substantive discussion ($\approx6\%$ lower odds). \emph{PC8} ($OR=0.93, q < 0.001$) shows that heavily time-stamped or event-dated contributions are penalized ($\approx7\%$ lower odds), and \emph{PC10} ($OR=0.95, q < 0.05$) suggests that serendipitous ``found this'' posts lack more effortful, curated, or reflective insight ($\approx5\%$ lower odds). 

The remaining PCs do not meet the BH–FDR correction threshold, and we therefore interpret them as \emph{directional} tendencies as observed from our data, rather than firm effects. We find that text-only discussions may fare slightly worse than visual and creative showcases (\emph{PC2}, $OR=0.99$). Furthermore, dense, list-like expositions (\emph{PC4}, $OR=0.98$) and off-topic/meta commentary (\emph{PC5}, $OR=0.98$) lean negative, lowering odds of receiving a high score by $\approx2\%$. Along similar lines, purely literal recounting has lower odds than reflective commentary (\emph{PC9}, $OR=0.97$) by $\approx3\%$.

On the other hand, encouragingly, creative and achievement oriented posts are equally well-received as personal struggle narratives on average (\emph{PC3}, $OR=1.00$). Lastly, enthusiastic, community-oriented calls lean positive (\emph{PC7}, $OR=1.03$), increasing odds of a high score by $\approx3\%$.

These semantic dimensions suggest a broader narrative about how communities distribute reward. High-scoring posts are not merely those that convey more or dense information, but those that frame contributions as invitations to shared sense-making. This could be through creativity, reflection, enthusiasm, or broader open-ended prompts. Conversely, styles that constrain dialogue, impose technical density, or signal low-effort contributions tend to be discounted by the community in the form of low upvotes.

Now that we have an understanding of the semantic styles that capture higher-level discourse orientations, we move to discuss surface linguistic markers, LIWC features, and toxicity.

\subsubsection{Surface-level Linguistic Features} We next examine surface-level linguistic markers, focusing on directly measurable properties of text such as interrogative usage, readability, and sentiment tone. Among the set of features we tested, two emerged as statistically significant predictors of whether a post achieves top–quartile score.

\textbf{Findings:} First, posts with a higher proportion of interrogative sentences are substantially disadvantaged, with an odds ratio $OR=0.71~(q < 0.001)$ indicating that question-heavy content is almost $30\%$ less likely to be rewarded. This broadly seems to echo the semantic findings around PC1 that narrow help-seeking requests tend to underperform. Second, readability exerts a consistent positive effect, with each one-point increase in the Flesch reading ease score translating to an odds ratio $OR=1.40~(q < 0.005)$. This highlights that clearer, more accessible posts are systematically preferred by users and rewarded using upvotes. 

While sentiment polarity did not yield a significant coefficient, we observe a counterintuitive pattern in terms of directionality, i.e., more positive sentiment slightly decreases the odds of high scores ($OR=0.98$), suggesting that content perceived as overly cheerful or one-dimensional may be slightly less compelling than posts with nuance, authenticity, or even controversy. 

In combination, these results highlight that some surface-level features indeed matter in shaping community response, and often do so in a way that reinforces broader stylistic discourse preferences. Communities seem to reward posts that are declarative rather than interrogative, accessible rather than complex structured, and authentic or perhaps even controversial rather than uniformly positive posts, although the latter is not causally relevant. 

\subsubsection{Toxicity} 

We find that toxicity is a statistically significant negative predictor of community reward. With an odds ratio $OR=0.95~(q < 0.001)$, thus indicating that an increase in toxicity score as measured by our toxicity model reduces the likelihood of a post reaching the top quartile by nearly $5\%$. This effect holds even after controlling for semantic style and other surface-level linguistic features, underscoring that toxic content is systematically disfavored. The finding highlights how communities across Reddit seem to penalize hostility, aggression, or derogatory tone, reinforcing a preference for constructive and respectful contributions as the foundation for positive feedback.

\subsubsection{LIWC Attributes}  

\begin{table}[t]
\sffamily
\small
\centering
\caption{LIWC categories causally associated with top–quartile score across 100 subreddits. Cells show odds ratios ($=\exp(\beta_k)$) from the ``high-score'' model. Odds ratios above 1 indicate dimensions that increase the likelihood of community reward, while ratios below 1 indicate decreased likelihood. Umbrella residuals (\emph{other\_}U) capture umbrella-level signal beyond their children. Significance levels for $\beta_k$: \textbf{***}\,$q<0.001$, \textbf{**}\,$q<0.01$, \textbf{*}\,$q<0.05$.}\label{tab:liwc}
\Description{Table with 22 rows organized into 8 thematic sections showing LIWC linguistic categories and their causal effects on Reddit post scores. Odds ratios range from 0.92 to 1.05, with 14 categories showing positive effects above 1.0 and 8 showing negative effects below 1.0. The strongest positive effect is adverbs at 1.05, while the strongest negative effect is conjunctions at 0.92. Statistical significance varies from single to triple asterisks, with most entries showing at least single asterisk significance. Color coding distinguishes positive effects in green background and negative effects in red background. The Informal language section shows mixed results with assent being positive and informal residual being negative. Linguistic dimensions section shows conjunctions as negative but adverbs as positive. Grammar categories including comparisons and numbers both show positive effects. Affective processes show all positive effects including positive emotion, anxiety, and sadness. Cognitive processes show mixed results with causation and cognitive residual positive but tentative negative. Perceptual processes split with see being negative and feel being positive. Biological processes show both bio residual and ingestion as negative. Drives residual shows positive effect, family mentions show positive effect, and future focus shows positive effect.}

\begin{tabular}{l S[table-format=2.2] l}
\rowcolor{gray!15}
\textbf{LIWC Category} & \textbf{Odds Ratio} & \textbf{$q$-sig} \\
\cmidrule(lr){1-1} \cmidrule(lr){2-2} \cmidrule(lr){3-3}
\addlinespace
\rowcolor{blue!10}\multicolumn{3}{c}{\textbf{Informal language}} \\
\emph{informal residual} (other\_informal) & \cellcolor{Crimson!15}\textbf{0.98} & \textbf{**} \\
Assent & \cellcolor{ForestGreen!15}\textbf{1.02} & \textbf{***} \\
\addlinespace
\rowcolor{blue!10}\multicolumn{3}{c}{\textbf{Linguistic Dimensions}} \\
Conjunctions & \cellcolor{Crimson!15}\textbf{0.92} & \textbf{***} \\
Adverbs & \cellcolor{ForestGreen!15}\textbf{1.05} & \textbf{***} \\
\addlinespace
\rowcolor{blue!10}\multicolumn{3}{c}{\textbf{Other Grammar}} \\
Comparisons & \cellcolor{ForestGreen!15}\textbf{1.03} & \textbf{***} \\
Numbers & \cellcolor{ForestGreen!15}\textbf{1.02} & \textbf{*} \\
\addlinespace
\rowcolor{blue!10}\multicolumn{3}{c}{\textbf{Affective Processes}} \\
Positive Emotion & \cellcolor{ForestGreen!15}\textbf{1.04} & \textbf{**} \\
Anxiety & \cellcolor{ForestGreen!15}\textbf{1.01} & \textbf{*} \\
Sadness & \cellcolor{ForestGreen!15}\textbf{1.01} & \textbf{**} \\
\addlinespace
\rowcolor{blue!10}\multicolumn{3}{c}{\textbf{Cognitive Processes}} \\
\emph{cogproc residual} (other\_cogproc) & \cellcolor{ForestGreen!15}\textbf{1.03} & \textbf{***} \\
Causation & \cellcolor{ForestGreen!15}\textbf{1.02} & \textbf{**} \\
Tentative & \cellcolor{Crimson!15}\textbf{0.97} & \textbf{*} \\
\addlinespace
\rowcolor{blue!10}\multicolumn{3}{c}{\textbf{Perceptual Processes}} \\
See & \cellcolor{Crimson!15}\textbf{0.98} & \textbf{***} \\
Feel & \cellcolor{ForestGreen!15}\textbf{1.01} & \textbf{***} \\
\addlinespace
\rowcolor{blue!10}\multicolumn{3}{c}{\textbf{Biological Processes}} \\
\emph{bio residual} (other\_bio) & \cellcolor{Crimson!15}\textbf{0.98} & \textbf{*} \\
Ingestion & \cellcolor{Crimson!15}\textbf{0.99} & \textbf{**} \\
\addlinespace
\rowcolor{blue!10}\multicolumn{3}{c}{\textbf{Drives}} \\
\emph{drives residual} (other\_drives) & \cellcolor{ForestGreen!15}\textbf{1.03} & \textbf{**} \\
\addlinespace
\rowcolor{blue!10}\multicolumn{3}{c}{\textbf{Social Processes}} \\
Family & \cellcolor{ForestGreen!15}\textbf{1.01} & \textbf{**} \\
\addlinespace
\rowcolor{blue!10}\multicolumn{3}{c}{\textbf{Time Orientations}} \\
Future focus & \cellcolor{ForestGreen!15}\textbf{1.02} & \textbf{*} \\
\bottomrule
\end{tabular}
\end{table}

In addition to semantic style dimensions and surface-level markers, we also examined the role of psycholinguistic attributes captured by the LIWC (Linguistic Inquiry and Word Count) lexicon. LIWC categories provide interpretable mappings between word use and psychological, social, and stylistic processes, allowing us to test whether systematic variation in these features predicts the probability of positive community feedback. Table~\ref{tab:liwc} reports the odds ratios from logit link linear model fit for the ``high score'' outcome, with several categories showing statistically significant and often large effects.

\textbf{Findings:} We now dive deeper into each category and our findings regarding the same. Due to the large number of LIWC categories, we only report findings and interpretations for the significant attributes here.

\emph{Informal language:} Posts with higher scores on the \emph{informal language residual} are slightly less likely to be rewarded ($OR=0.98, q < 0.01$). In contrast, a specific conversational cue, \emph{assent} (e.g., \emph{``agree,'' ``OK''}) within this category is positively associated with success ($OR=1.02, q<0.001$). As stated before, when we model the umbrella category alongside its subcategories, the umbrella coefficient reflects residual, diffuse informality that is not captured by the targeted children categories. The pattern implies that light markers of agreement (i.e., subcategory assent) can signal warmth and alignment, whereas broad informality that adds noise or reduces clarity  (i.e., the informal language residual) tends to be down-weighted.

\emph{Linguistic dimensions:} We again observe a split pattern in grammatical scaffolding. \emph{Conjunctions} (e.g., \emph{``and,'' ``whereas''}) correspond to lower odds ($OR=0.92$) by $\approx8\%$, consistent with the penalty for more tangled sentence linking that we saw in the stylistic PC dimensions. By contrast, \emph{adverbs} (e.g., \emph{``very,'' ``really''}) are associated with higher odds ($OR=1.05$) by $\approx5\%$. These results suggest that heavy connective chaining can make posts feel verbose, while the use of adverbs can add emphasis on the users contributions without harming readability.

\emph{Other grammar:} With more general grammar, we find that lexical choices that add precision to the poster's contributions continue to help in increasing odds of receiving a high score. \emph{Comparisons} (e.g., \emph{``greater,'' ``best''}) predict higher success ($OR=1.03$) with $\approx3\%$ high odds, as do explicit \emph{numbers} (e.g., \emph{``second,'' ``thousand''}; $OR=1.02$). We therefore find that specificity and comparative framing aid comprehension and credibility, which aligns with the broader association we see between clarity and positive feedback.

\emph{Psychological processes:} With psychological processes different categories of processes show different trends.
\begin{itemize}\setlength{\itemsep}{0pt}\setlength{\parskip}{0pt}
    \item \emph{Affective processes:} These markers capture expressed emotion within the post. We see modest positive associations for emotional expression. \emph{Positive emotion} terms (e.g., \emph{``happy,'' ``love''}) correspond to higher odds ($OR=1.04$). Even \emph{anxiety} words (e.g., \emph{``worried,'' ``nervous''}; $OR=1.01$) and \emph{sadness} words (e.g., \emph{``sad,'' ``crying''}; $OR=1.01$) show small positive links. This pattern indicates that visible feeling and emotion can aid reception when expressed in a grounded or empathetic way, while our separate toxicity features likely absorb the effect of hostile tone that would otherwise depress outcomes.
    
    \item \emph{Cognitive processes:} These terms indicate reasoning and stance-taking within the post. The \emph{cognitive processes residual} is positively associated with success ($OR=1.03$), and \emph{causation} language (e.g., \emph{``because,'' ``effect''}) also shows a positive association ($OR=1.02$). In contrast, \emph{tentative} markers (e.g., \emph{``maybe,'' ``perhaps''}) predict lower odds ($OR=0.97$). In practice, readers reward clear causal framing and overall reasoning, while excessive hedging still dampens perceived confidence and impact.
    
    \item \emph{Perceptual processes:} These terms ground ideas in sensory or felt experience. \emph{Feel} markers (e.g., \emph{``feel,'' ``touch''}) are positively associated with scoring ($OR=1.01$), indicating benefits from embodied or experiential references. Purely visual mentions captured by \emph{see} terms (e.g., \emph{``look,'' ``see''}) show a slight penalty ($OR=0.98$), which suggests that pointing to experience works best when it conveys affect or sensation rather than simple visual noting.
    
    \item \emph{Biological processes:} These markers reference bodily states and needs. The \emph{biological residual} is associated with lower odds ($OR=0.98$), and \emph{ingestion} terms (e.g., \emph{``eat,'' ``drink''}) also show a small penalty ($OR=0.99$). We hypothesize that concrete body or eating talk may feel off-topic in many communities, which nudges reception downward.
    
    \item \emph{Drives:} Drive words point to motivations and goals expressed within the post. The \emph{drives residual} is positively associated with success ($OR=1.03$). General motivational tone appears to be appreciated when it reads as authentic, and perhaps builds a sense of shared goal-setting.
    
    \item \emph{Social processes:} These markers reference people and relationships, that is, who the post involves. We find that concrete mentions of relational anchors help. \emph{Family} mentions (e.g., \emph{``mom,'' ``dad''}) predict higher odds ($OR=1.01$) by $\sim1\%$. Personal references to specific roles can perhaps create a shared social presence.
    
    \item \emph{Time orientations:} We also find temporal anchoring to be beneficial. We observe a positive association for \emph{future focus} (e.g., \emph{``will,'' ``soon''}; $OR=1.02$). Situating content \emph{prospectively} can signal intent and next steps, which supports engagement, also supporting our previous result on \emph{drives}.
\end{itemize}

Therefore across psychological processes, readers appear to favor posts that are reflective and grounded, with cognitive language used to explain causes, experiential anchoring that feels relatable, and a clear temporal frame that is forward-looking. At the same time, styles that feel hedged without payoff, broadly informal, or includes mentions of biological processes in a way that drifts off topic tend to underperform.

\subsubsection{Topics:} \edit{Across varying number of topics ($K$) in our LDA-based topic modeling, we found no statistically significant predictors of high score in the set of topic features we modeled.} This indicates that no particular topic was unilaterally rewarded across all community using upvotes in our observation window, which might be expected since the communities we study are very diverse in topics\edit{, which may not be adequately captured with $5-25$ topics.}

\textbf{Takeaway:} Overall, these findings highlight that communities across Reddit positively reward contributions with \emph{clear reasoning with minimal hedging}, \emph{concrete personal or relational anchors}, and \emph{specific and comparative language that sharpens claims}. Light conversational alignment cues such as assent help as well. Negativity that strays into hostility in the form of toxicity is disfavored, although authentic emotional expression of angst or sadness can coexist with strong reception. 

\medskip

\textbf{Note:} To assess temporal stability of our findings, we split our observation window into two non-overlapping periods (May-July 2020 and July-September 2020) and  re-estimated all models independently. The correlation between log-odds ratios across periods is $r = 0.764 ~(p < 0.001)$, indicating our findings reflect stable patterns rather than period-specific artifacts (see \autoref{app:temporal-robustness} for details).

\begin{table*}[t]
\centering
\small
\sffamily
\caption{Predictors of \textit{awards} across 100 subreddits. Odds ratios ($=\exp(\beta_k)$) are estimated from a logistic model fit for the ``award'' outcome. Odds ratios above 1 indicate dimensions that increase the likelihood of community reward, while ratios below 1 indicate decreased likelihood. Arrows (↑↑ = increases odds, ↓↓ = decreases odds) summarize practical direction of implications for communities. Significance levels for $\beta_k$: \textbf{***}\,$q<0.001$, \textbf{**}\,$q<0.01$, \textbf{*}\,$q<0.05$.}
\label{tab:awards}
\Description{Table with 12 rows organized into three thematic sections showing predictors of Reddit awards. Odds ratios range from 0.81 to 1.15, with seven predictors showing positive effects above 1.0 and five showing negative effects below 1.0. The strongest positive effect is PC1 at 1.15 for broad engagement prompts, while the strongest negative effect is question ratio at 0.81. Statistical significance varies across predictors, with PC1, PC4, question ratio, and money showing triple asterisk significance. The Stylistic Dimensions section shows mixed results with PC1, PC3, and PC7 positive but PC4 and PC8 negative. Surface-level features split with question ratio strongly negative at 0.81 and VADER sentiment slightly positive at 1.04. LIWC attributes show mostly negative effects for conjunctions, adverbs, and tentative language, but money mentions show positive effects at 1.03. Color coding distinguishes positive effects in green background and negative effects in red background. Directional arrows in the rightmost column indicate practical implications with double up arrows for strong positive effects and double down arrows for strong negative effects.}

\begin{tabular}{lcccl}
\rowcolor{gray!15}
\textbf{Predictor} & \textbf{Odds Ratio} & \textbf{$q$-sig} & \textbf{Interpretation from exemplars} & \textbf{Implication for awards} \\
\cmidrule(lr){1-1}\cmidrule(lr){2-2}\cmidrule(lr){3-3}\cmidrule(lr){4-4}\cmidrule(lr){5-5}
\addlinespace
\rowcolor{blue!10}\multicolumn{5}{c}{\textbf{Stylistic Dimensions}} \\
PC1 & \cellcolor{ForestGreen!15}\textbf{1.15} & \textbf{***} & Broad engagement prompts over narrow help & \cellcolor{ForestGreen!15}↑↑ Engagement framing rewarded \\
PC3 & \cellcolor{ForestGreen!15}\textbf{1.05} & \textbf{*} & Creative or achievement narratives & \cellcolor{ForestGreen!15}↑↑ Slight boost for showcases \\
PC4 & \cellcolor{Crimson!15}\textbf{0.93} & \textbf{***} & Dense list-like exposition & \cellcolor{Crimson!15}↓↓ Dense enumerations penalized \\
PC7 & \cellcolor{ForestGreen!15}\textbf{1.05} & \textbf{*} & Enthusiastic, community-oriented calls & \cellcolor{ForestGreen!15}↑ Collective enthusiasm helps \\
PC8 & \cellcolor{Crimson!15}\textbf{0.97} & \textbf{**} & Event-dated or time-stamped content & \cellcolor{Crimson!15}↓↓ Heavily time-bound posts penalized \\
\addlinespace
\rowcolor{blue!10}\multicolumn{5}{c}{\textbf{Surface-level Linguistic Features}} \\
Question ratio & \cellcolor{Crimson!15}\textbf{0.81} & \textbf{***} & Question-heavy framing & \cellcolor{Crimson!15}↓↓ Fewer awards for questions \\
VADER sentiment & \cellcolor{ForestGreen!15}\textbf{1.04} & \textbf{*} & Overall valence & \cellcolor{ForestGreen!15}↑↑ Positive tone slightly rewarded \\
\addlinespace
\rowcolor{blue!10}\multicolumn{5}{c}{\textbf{LIWC Attributes}} \\
Conjunctions & \cellcolor{Crimson!15}\textbf{0.97} & \textbf{**} & Connectives (e.g., ``and,'' ``whereas'') & \cellcolor{Crimson!15}↓↓ Over-connected prose penalized \\
Adverbs & \cellcolor{Crimson!15}\textbf{0.98} & \textbf{*} & Degree and manner modifiers & \cellcolor{Crimson!15}↓↓ Measured descriptiveness fares worse \\
Tentative & \cellcolor{Crimson!15}\textbf{0.98} & \textbf{*} & Hedging (e.g., ``maybe,'' ``perhaps'') & \cellcolor{Crimson!15}↓↓ Hedging penalized \\
Money & \cellcolor{ForestGreen!15}\textbf{1.03} & \textbf{***} & Mentions of money or resources & \cellcolor{ForestGreen!15}↑↑ Resource/benefit posts rewarded \\
\bottomrule
\end{tabular}
\end{table*}

\subsection{RQ1(b): Do drivers of positive reward differ across feedback outcomes?}

Extending our analysis on high score, we model \textit{awards} and \textit{gold} using the same covariates, preprocessing, and fixed effects as in RQ1(a). Furthermore, principal components orientations match \autoref{tab:pc-dimensions} since PCs were computed prior to modeling. Below we report main effects only, with the discussion on interaction effects postponed to RQ1(c).

\textbf{Findings:} Awards share several drivers with \textit{score} but not all. Engagement-oriented framing continues to be favored (PC1; $OR=1.15, q<0.001$). Enthusiastic, community-oriented calls show a small positive association (PC7; $OR=1.05, q<0.05$), and creative or achievement showcases are modestly rewarded (PC3; $OR=1.05, q<0.05$). However, several styles are disfavored, such as question-heavy framing carrying a clear penalty with $OR=0.81, q<0.001$, reducing odds of receiving an award by almost $20\%$. Similarly, dense list-like exposition underperforms (PC4; $OR=0.93, q<0.001$), and heavily time-bound posts are also less well-received (PC8; $OR=0.97, q<0.01$). Once again, we find that overly complex prose is penalized with \emph{conjunctions} lowering odds of an award by $\approx3\%$ ($OR=0.97, q<0.01$), as is hedging (\emph{tentative}; $OR=0.98, q<0.05$). Furthermore, adverbs weigh negatively on the likelihood of receiving \textit{awards} ($OR=0.98, q<0.05$), while a slightly more positive overall sentiment is now beneficial (VADER; OR$=1.04, q<0.05$). Mentions of money or resources are also positively associated with awards (\emph{money}; $OR=1.03, q<0.001$).

\begin{table}[t]
\small
\sffamily
\centering
\caption{Predictors of \textit{golds} across 100 subreddits. Odds ratios ($=\exp(\beta_k)$) are estimated from a logistic model fit for the ``award'' outcome. Odds ratios above 1 indicate dimensions that increase the likelihood of community reward, while ratios below 1 indicate decreased likelihood. Arrows (↑↑ = increases odds, ↓↓ = decreases odds) summarize practical direction of implications for communities. Significance levels for $\beta_k$: \textbf{***}\,$q<0.001$, \textbf{**}\,$q<0.01$, \textbf{*}\,$q<0.05$.}
\label{tab:gold}
\Description{Simple three-row table showing LIWC linguistic predictors of Reddit gold awards. All three predictors show single asterisk statistical significance. Odds ratios range from 0.94 to 1.06, indicating relatively modest effects compared to other feedback outcomes. Power language shows negative association at 0.94 with red background coding, while sadness and money mentions both show positive associations at 1.06 and 1.04 respectively with green background coding. The rightmost column uses double directional arrows to indicate practical implications, with power language disfavored and both sadness expressions and resource mentions being rewarded. This represents the most restrictive set of significant predictors among the three feedback outcomes, with only three categories reaching statistical significance compared to the larger sets for high scores and awards.}

\resizebox{\columnwidth}{!}{
\begin{tabular}{lccl}
\rowcolor{gray!15}
\textbf{Predictor} & \textbf{Odds Ratio} & \textbf{$q$-sig} & \textbf{Implication for gold} \\
\cmidrule(lr){1-1}\cmidrule(lr){2-2}\cmidrule(lr){3-3}\cmidrule(lr){4-4}
\addlinespace
\rowcolor{blue!10}\multicolumn{4}{c}{\textbf{LIWC Attributes}} \\
Power & \cellcolor{Crimson!15}\textbf{0.94} & \textbf{*} & \cellcolor{Crimson!15}↓↓ Power language disfavored \\
Sadness & \cellcolor{ForestGreen!15}\textbf{1.06} & \textbf{*} & \cellcolor{ForestGreen!15}↑↑ Empathic responses to sadness \\
Money & \cellcolor{ForestGreen!15}\textbf{1.04} & \textbf{*}  & \cellcolor{ForestGreen!15}↑↑ Resource / benefit posts rewarded \\
\bottomrule
\end{tabular}}
\end{table}

\textit{Gold} is rarer and yields a narrower set of predictors. From \autoref{tab:gold}, we see that power and status-focused language is disfavored ($OR=0.94, q<0.05$), while expressed sadness is positively associated ($OR=1.06, q<0.05$). Similar to awards, mentions of money or resources also show a modest positive association with being gilded ($OR=1.04, q<0.05$). 

\subsubsection{Agreement and differences with predictors for ``score'':} Across channels we see overlaps but also clear divergences. Engagement-oriented framing aligns for both \emph{score} and awards, with broad, discussion-generating posts being favored in both cases. Similarly, heavily time-bound content and connective-heavy prose is penalized for both outcomes. However, several drivers differ from \emph{score}. For example the use of adverbs flips its association from score to awards, being positive for \emph{score} but slightly negative for awards. A slightly more positive overall sentiment matters for awards while it is not a robust causal driver for \emph{score}. Furthermore, resource-heavy framing shows up in monetarily discretionary channels, with mentions of money being positively associated with both awards and Gold, but are not a key predictor for \emph{score}. Finally, Gold yields the narrowest set of predictors across all channels, showing a penalty for power and status-driven language and a modest lift for expressions of sadness. \edit{We view this heterogeneity as a substantive finding and strength of our work. It reveals that free voting mechanisms and paid endorsements encode distinct forms of community value---a distinction obscured by prior surveys that treat all positive feedback as largely equivalent~\cite{lambert_positive_2024,weld_making_2024}.}

\subsubsection{Why we find fewer significant predictors:} As evident from these results, we find much fewer significant predictors that causally lower or higher the likelihood of a contribution receiving an award or gold, in comparison to upvotes. We hypothesize a few reasons for this. First, awards and gold are sparsely used on the platform, which lowers statistical power and makes moderate effects harder to detect after multiple testing correction. Second, these feedback channels involve monetary payments, which adds selectivity in a way that the set of gilded or awarded posts likely reflects specific interests and social signaling rather than broad community preferences. Finally, we believe that non-linguistic factors such as attached media, prior visibility from cross posts, and community specific norms could also play a larger role for these outcomes. These drivers are only partly captured by fixed effects and lie outside the scope of the linguistic and semantic features studied in our work.

\subsection{RQ1(c): How do feature effects vary with author status (newcomers vs. veterans)?}

We test whether linguistic effects differ for newcomers (accounts $<90$ days old) by adding interaction terms of the form $A_k \times \textsc{NEW}$ to the score model. The \emph{main} $\textsc{NEW}$ coefficient captures a baseline difference in odds for newcomers at the mean of $A_k$ (features are $z$–scored). An interaction coefficient scales the feature effect for newcomers, such that
\begin{align}
    \mathrm{OR}_{\text{new}}(A_k) \;=\; \mathrm{OR}_{\text{vet}}(A_k)\times \mathrm{OR}_{A_k\times \textsc{NEW}}.
\end{align}

\subsubsection{High score outcome:} Upon BH-FDR correction ($q < 0.05$), we found three significant interactions for \emph{score}.

\begin{table}[t]
\small
\sffamily
\centering
\caption{Newcomer interactions for the ``\emph{high score}'' outcome across 100 subreddits. ``Interaction OR'' multiplies the veteran effect to yield the newcomer effect. Significance levels for $\beta_k$: \textbf{***}\,$q<0.001$, \textbf{**}\,$q<0.01$, \textbf{*}\,$q<0.05$. ORs not significant are marked with $-$.}
\label{tab:newcomer-interactions-score}
\Description{Table with three rows showing interaction effects between newcomer status and linguistic features for high score outcomes. The table has five columns: feature name, interaction odds ratio, statistical significance, veterans odds ratio, and newcomers odds ratio. Readability shows the strongest and most complete effect with interaction odds ratio of 1.18, veterans odds ratio of 1.40, and newcomers odds ratio of 1.65, all with green background coding indicating positive effects. Netspeak and Negation rows show interaction odds ratios of 0.98 and 1.02 respectively, but have dashes in the veterans and newcomers columns indicating non-significant main effects. All three features show single asterisk statistical significance. The readability result demonstrates that while clear writing benefits both groups, newcomers receive disproportionately larger benefits from readable content compared to veteran users.}
\resizebox{\columnwidth}{!}{
\begin{tabular}{lcccc}
\rowcolor{gray!15}
\textbf{Feature} & \textbf{Interaction OR} & \textbf{$q$-sig} & \textbf{Veterans OR} & \textbf{Newcomers OR} \\
\cmidrule(lr){1-1}\cmidrule(lr){2-2}\cmidrule(lr){3-3}\cmidrule(lr){4-4}\cmidrule(lr){5-5}
Readability (Flesch) & \cellcolor{ForestGreen!15}\textbf{1.18} & \textbf{*} & \cellcolor{ForestGreen!15}\textbf{1.40} & \cellcolor{ForestGreen!15}\textbf{1.65} \\
Netspeak & \cellcolor{Crimson!15}\textbf{0.98} & \textbf{*} & — & — \\
Negation & \cellcolor{ForestGreen!15}\textbf{1.02} & \textbf{*} & — & — \\
\bottomrule
\end{tabular}}
\end{table}

\textbf{Findings (score):} We firstly find that newcomers face a small but significant baseline disadvantage even after controls and fixed effects ($\textsc{NEW}$ $OR=0.94$, $q<0.01$), indicating $\approx6\%$ lower odds of reaching top–quartile score holding observed factors constant compared to a veteran. Next, from \autoref{tab:newcomer-interactions-score} we find that readability benefits newcomers more than veterans. While veterans gain $\approx$40\% higher odds per SD increase in Flesch readability score, the interaction further lifts newcomers to $\approx$65\% higher odds per SD increase. Netspeak (e.g., \emph{``btw,'' ``lol''}) is slightly less effective for newcomers than for veterans (interaction $OR=0.98, q<0.05$). Negation words (e.g., \emph{``not,'' ``never''}) shows a small positive moderation (interaction $OR=1.02, q<0.05$). While these interactions are small in size and the veteran and newcomer OR are not statistically different from $1$, newcomer advantage for readability is practically meaningful.

\subsubsection{Award and gold outcomes:}
For \textit{gold}, we do not detect significant newcomer interactions. However, for \textit{awards}, we find three interactions that surpass the BH--FDR threshold ($q<0.05$):

\begin{table}[t]
\small
\sffamily
\centering
\caption{Newcomer interactions for the \emph{``awards''} outcome. Interaction OR $>1$ indicates the feature is more beneficial for newcomers \emph{relative to} veterans; $<1$ indicates less beneficial. ORs not significant are marked with $-$.}
\label{tab:newcomer-interactions-award}
\Description{Table with three rows showing interaction effects between newcomer status and linguistic features for award outcomes. The table displays five columns showing feature names, interaction odds ratios, statistical significance, veterans odds ratios, and newcomers odds ratios. Future focus shows positive interaction at 1.03 with double asterisk significance, benefiting both veterans at 1.02 and newcomers at 1.05, with green background coding. Anxiety shows negative interaction at 0.97 with single asterisk significance, where veterans benefit at 1.01 but newcomers are penalized at 0.98, creating a mixed color pattern. Informal residual shows negative interaction at 0.97 with single asterisk significance, penalizing both veterans at 0.98 and newcomers more severely at 0.95, both with red background coding. The pattern demonstrates that future-focused language benefits newcomers more than veterans, while anxiety expressions and informal language create differential effects, with newcomers facing greater penalties for informal language and receiving less benefit from anxiety expressions compared to veteran users.}
\resizebox{\columnwidth}{!}{
\begin{tabular}{lcccc}
\rowcolor{gray!15}
\textbf{Feature} & \textbf{Interaction OR} & \textbf{$q$-sig} & \textbf{Veterans OR} & \textbf{Newcomers OR} \\
\cmidrule(lr){1-1}\cmidrule(lr){2-2}\cmidrule(lr){3-3}\cmidrule(lr){4-4}\cmidrule(lr){5-5}
Future focus & \cellcolor{ForestGreen!15}\textbf{1.03} & \textbf{**} & \cellcolor{ForestGreen!15}\textbf{1.02} & \cellcolor{ForestGreen!15}\textbf{1.05} \\
Anxiety & \cellcolor{Crimson!15}\textbf{0.97} & \textbf{*} & \cellcolor{ForestGreen!15}\textbf{1.01} & \cellcolor{Crimson!15}\textbf{0.98} \\
\emph{informal residual} & \cellcolor{Crimson!15}\textbf{0.97} & \textbf{*} & \cellcolor{Crimson!15}\textbf{0.98} & \cellcolor{Crimson!15}\textbf{0.95} \\
\bottomrule
\end{tabular}}
\end{table}

\textbf{Findings (awards):} As summarized in \autoref{tab:newcomer-interactions-award}, some newcomer interactions for \textit{awards} are small in magnitude but statistically reliable after BH--FDR. \emph{Future focus} is beneficial for both groups and more so for newcomers (Veterans $OR=1.02$, Newcomers $OR=1.05$, Interaction $OR=1.03, q<0.01$ per 1 SD increase), suggesting that prospective framing helps new accounts invite more participation and positive feedback to their post. In contrast, \emph{anxiety} is slightly positively received for veterans but penalized for newcomers (Veterans $OR=1.01$ \emph{vs.} Newcomers $OR=0.98$, Interaction $OR=0.97, q<0.05$). Moreover, the \emph{informal residual} (\emph{other\_informal}) as we saw is disfavored for veterans, but even more strongly so for newcomers (Veterans $OR=0.98$, Newcomers $OR=0.95$, Interaction $OR=0.97, q<0.05$).

\textbf{Takeaway:} Overall, these findings reveal that newcomers at a baseline level are already disadvantaged compared to veterans with $\approx6\%$ lower odds of receiving a high score, keeping all other attributes constant. Across outcomes, newcomers can partially substitute for limited reputation with \emph{clarity} and \emph{prospective framing}, whereas community-specific informal tone and anxiety-loaded language fare less well for new accounts.

\begin{figure*}
    \centering
    \includegraphics[width=\linewidth]{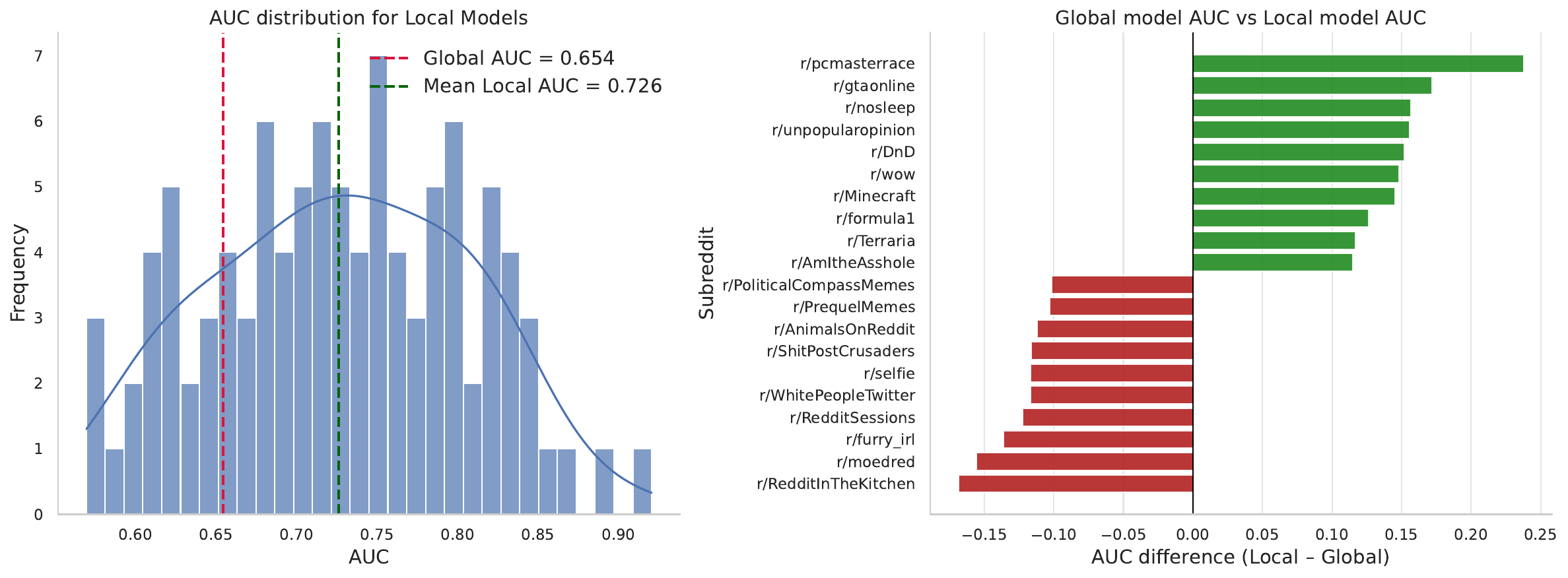}
    \caption{\textbf{(Left)} Distribution of AUC scores for subreddit‑specific (local) models trained on posts from specific subreddits; the red dashed line marks the global model’s overall AUC (0.654), while the green dashed line marks the average AUC of local models (0.726). \textbf{(Right)} Per‑subreddit AUC difference between local and global models ($\Delta=$ Local AUC - Global AUC) evaluated on the test set of each subreddit: green bars to the right of $0$ indicate local models outperform the global model, red bars to the left of $0$ indicate the global model performs better. Only the top-10 gains and losses are shown.}
    \Description{Left panel shows histogram of Area Under the Curve scores ranging from 0.60 to 0.90 with frequency on y-axis up to 7. Distribution peaks around 0.72 to 0.74 with bell-shaped curve overlaid. Two vertical reference lines mark global model performance at 0.654 and mean local performance at 0.726. Right panel displays horizontal bar chart with 20 subreddit names listed vertically, showing performance differences ranging from negative 0.15 to positive 0.25. Top performers include pcmasterrace, gtaonline, nosleep, unpopularopinion, and DnD with substantial positive differences. Bottom performers include PrequelMemes, AnimalsOnReddit, ShitPostCrusaders, selfie, and RedditInTheKitchen with negative differences, indicating cases where global models outperform local ones.}
    \label{fig:local-global-auc}
\end{figure*}

\section{RQ2: Predictive Modeling for Proactive Identification of Desirable Comments}

We now ask whether the pre-feedback signals studied in \emph{RQ1} can support real-time discovery of high quality posts, that is, whether models trained on features observable at posting time can surface promising content that may be underrepresented while votes are still organically accruing. Since the strongest and most stable effects in \emph{RQ1} were for the \emph{score} outcome, we focus on predictive modeling \edit{only for \emph{score} to focus on the most important outcome.}

\subsection{Models and Metrics} We train two complementary style of detectors that turn the \emph{RQ1} feature set into working predictors. A \emph{global} model is fit on the pooled data from all 100 subreddits with a stratified split so that each community is represented in both the train and test sets. In parallel, we fit \emph{local} models, one per subreddit, where splitting and training occur within community. We trained gradient-boosted decision trees (XGBoost~\cite{chen2016xgboost} with \texttt{max\_depth=6}) with an $80{:}20$ train–test split, the most popular choice of models for tabular data.

The set of predictive feature space mirrors \emph{RQ1}, with semantic principal components, surface markers such as readability and question ratio, LIWC attributes including umbrella residuals, and toxicity. We evaluate performance of models using Area Under the ROC Curve (AUC) rather than Accuracy, since our prediction task is class-imbalanced and the desired operating point during real-time deployment is \emph{ranking} rather than hard classification. AUC is therefore a threshold-free metric and captures ranking quality under imbalance.

\subsection{RQ2(a): How well can the linguistic features from RQ1 predict positive feedback?}

\autoref{fig:local-global-auc} summarizes performance of the trained models at the global and local levels. The left panel shows the distribution of test AUCs for local models, with the red dashed line marking the global model on the pooled test set. The global AUC is $0.654$. We find that local models perform better on average with a mean AUC of $0.726$, with most subreddit-specific models performing better than the global AUC, and several even achieving substantially higher performance going up to $0.91$. This indicates that in many cases, local models exploit community-specific regularities that a single global model could average away.

\subsubsection{How does our model compare to simple existing desirability measures?}

To get a deeper understanding of how well our models perform, we compare the AUC of our model to the prosociality models introduced by \citet{bao_conversations_2021}, which has been used in prior HCI and social computing work~\cite{goyal_uncovering_2024,lambert_conversational_2022} as a measure for ``desirable'' contributions. Moreover, prosociality is considered a key component of desirable contributions on Reddit by moderators~\cite{lambert_positive_2024}.

Specifically, we follow the same framework as \citet{goyal_uncovering_2024} to use the BERT models
trained by \citet{bao_conversations_2021} to assign each post a \emph{supportiveness}, \emph{agreement}, and \emph{politeness} score, and then normalized all scores. We then used these three scores to synthesized a single-dimensional prosociality measure using  Principal Component Analysis (PCA), capturing $\approx78\%$ of the variance in the original
data. We then regressed the binary ``high score'' label on the post's synthesized prosociality score.

To ensure a fairer comparison, we also re-run our own modeling using a logistic regression classifier. We find that the prosociality-based predictor achieves a global AUC of $0.573$, which represents a $10\%$ drop in AUC compared to our predictive modeling using the logistic regression classifier $\textsc{AUC}=0.638$, and a $12\%$ drop in AUC compared to the XGBoost classifier we described previously. This highlights that the linguistic features set we curate help to outperform existing models to detect potentially desirable comments in online communities on Reddit.

\subsection{RQ2(b): How does the performance of predictive models vary across communities?} 

The right panel plot in \autoref{fig:local-global-auc} shows the communities with the AUC differences (gains and loss) between the local model and the global model evaluated on that community's test set. \edit{Since our dataset contains 100 subreddits, to extract meaningful insights while keeping interpretations free of bias, we focus on subreddits with both the top-10 \emph{gains} and top-10 \emph{losses} as a case study.}

The largest gains (in green) appear in \emph{r/pcmasterrace}, \emph{r/gtaonline}, \emph{r/nosleep}, \emph{r/unpopularopinion}, \emph{r/DnD}, \emph{r/wow}, \emph{r/Minecraft}, \emph{r/formula1}, \emph{r/Terraria}, and \emph{r/AmItheAsshole}, with the local models performing between $0.10$ to $0.23$ better than the global model in terms of AUC. We hypothesize that these forums likely have distinctive topics or narrative styles, so local models can specialize and outperform a global model that averages across heterogeneous contexts. 

The largest losses (in red) on the other hand occurs in \emph{r/selfie}, \emph{r/PrequelMemes}, \emph{r/ShitPostCrusaders}, \emph{r/PoliticalCompassMemes}, \emph{r/mo\-edred}, \emph{r/AnimalsOnReddit}, \emph{r/WhitePeopleTwitter}, \emph{r/RedditInTheKitch\-en}, \emph{r/furry\_irl}, and \emph{r/RedditSessions}, with the local models lagging behind the global models by $0.10$ to $0.17$ in terms of AUC. We hypothesize that these subreddits are highly memetic or visual, so text features capture only part of the signal. In such cases the global model benefits from being trained on more data and capturing broader regularities of desirability across Reddit.

\subsubsection{Does distinctiveness play a role?} To explore this pattern further \edit{and identify ways to mitigate this disparity}, we test whether our distinctiveness hypothesis is true. To quantify how linguistically ``far'' a subreddit's language is from the Reddit‐wide norm, we work with the same 384-dimensional MiniLM embeddings~\cite{reimers_sentence-bert_2019} from our feature set $\mathbf A$.

\edit{Operationalizing distinctiveness directly is challenging as it could depend on a variety of factors. To tackle this, we take inspiration from the prior work of \citet{10.1145/3613904.3642769} to use a distance-based metric for distinctiveness. Specifically,} for every subreddit $s$ we take a randomly sampled set of $N_s=2,000$ (total $N$ across all subreddits) its posts and compute their embeddings \(\{\boldsymbol e_{s,i}\}_{i=1}^{N_s}\). We then compute 
\begin{align}
\boldsymbol{\mu}_s \,=\, \frac{1}{N_s}\sum_{i=1}^{N_s}\boldsymbol e_{s,i}, \qquad
\boldsymbol{\mu}_{\text{global}} \,=\, \frac{1}{N}\sum_{j=1}^{N}\boldsymbol e_{j}
\end{align}
which represent the subreddit-wise and global centroids of these embeddings, respectively.

We define the linguistic distinctiveness of subreddit $s$ as the cosine distance between the subreddit-specific centroid $\boldsymbol{\mu}_s$ and the global centroid $\boldsymbol{\mu}_{\text{global}}$, 
\begin{align}
    d_s \,=\, 1\;{-}\;\frac{\boldsymbol{\mu}_s^{\mathsf T}\boldsymbol{\mu}_{\text{global}}}{\|\boldsymbol{\mu}_s\|_2\,\|\boldsymbol{\mu}_{\text{global}}\|_2}
\end{align}
which ranges from $0$ (identical direction) to $2$ (opposite direction). This way, we obtain a scalar $d_s$ per community.

Let $\mathcal{G}$ be the set of ``gain'' subreddits (\emph{local} > \emph{global}) and $\mathcal{L}$ the ``loss'' subreddits (\emph{global} > \emph{local}).  We test whether the mean distance differs between the two groups using Welch’s unequal-variance $t$–test~\cite{welch1947generalization}:
\begin{align}
 t \,=\, \frac{\bar d_{\mathcal G} - \bar d_{\mathcal L}}{\sqrt{\tfrac{s^2_{\mathcal G}}{|\mathcal G|} \;{+}\; \tfrac{s^2_{\mathcal L}}{|\mathcal L|}}}
\end{align}
where \(\bar d_{\mathcal G}\) and \(\bar d_{\mathcal L}\) represent the means of each set.

Using the $10$ largest gains and losses we obtain \(\bar d_{\mathcal G} = 0.393\) and \(\bar d_{\mathcal L} = 0.257\) with $t \approx 2.4,\,\,p=0.029$, which is significant at $\alpha=0.05$.\footnote{We replicated this test with $N_s=10,000$ and arrive at the same conclusion with $t \approx 2.1,\,\,p=0.036$.}

This implies that gain subreddits are, on average, roughly 50\% further away from the global centroid than loss subreddits, and this finding is statistically significant. These results confirm our hypothesis that communities with more semantically distinctive discourse benefit from specialized (\emph{local}) models, whereas communities whose language is closer to average posts on Reddit, or whose distinctiveness lies mainly in lexical quirks rather than semantic content, are better served by the larger, global model. More concretely, semantically less distinctive communities like the meme- or image-centric communities do not furnish enough signal for a \emph{local} model, and the \emph{global} model's broader training set provides better generalization.

\textbf{Takeaway:} Together, \emph{RQ2} shows that the same pre-feedback linguistic attributes that exhibit causal links to score in \emph{RQ1} also carry strong predictive signal. A simple XGBoost detector built on these features achieves $0.654$ AUC in the pooled setting and $0.726$ mean AUC for community specific models, also outperforming existing approaches. We therefore highlight that we can meaningfully rank promising posts at the moment of creation, bridging our causal findings and their practical modeling utility.

\begin{figure*}[t]
    \centering
    \includegraphics[width=0.8\linewidth]{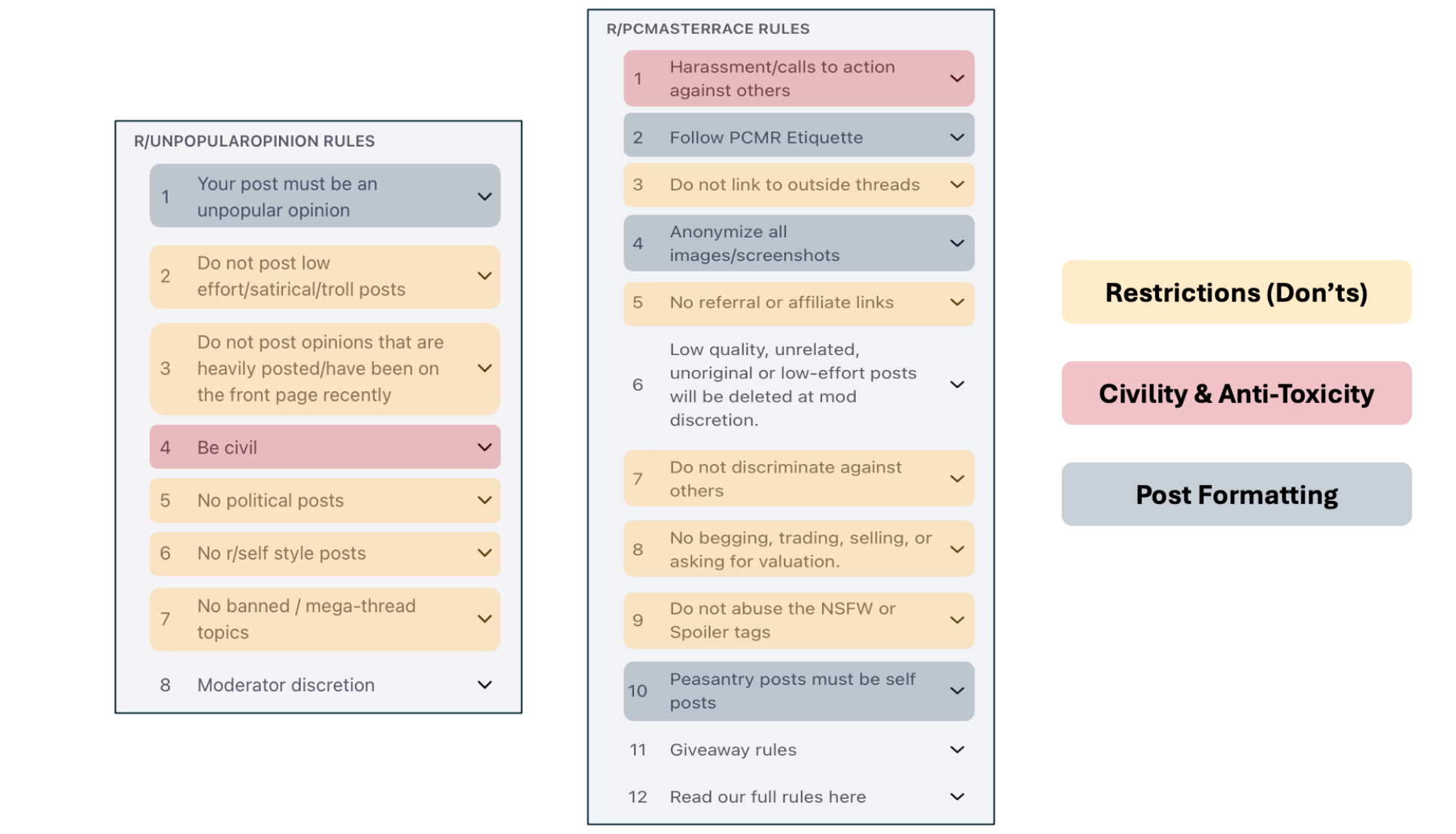}
    \caption{Color-coded rules for \textit{r/unpopularopinion} and \textit{r/pcmasterrace} subreddits. Most rules in each of the subreddits falls under the ``Restrictions'' category (yellow), highlighting what the user \textit{cannot} do. Other rules include anti-toxicity guidelines (red) or post-formatting guidelines (blue). None of the rules or their framing highlight how a user can make ``desirable'' contributions to teach them the values of the community.}
    \Description{Two side-by-side panels displaying subreddit rule interfaces with numbered dropdown menus. Left panel shows eight rules for unpopularopinion subreddit, right panel shows twelve rules for pcmasterrace subreddit. Rules are color-coded with yellow highlighting the majority of restrictions, red indicating civility and anti-toxicity guidelines, and blue marking post formatting requirements. The yellow restriction category dominates both panels, with approximately six out of eight rules in the left panel and eight out of twelve rules in the right panel falling into this category. Pink and blue categories each represent only one to two rules per subreddit. The visual demonstrates the predominance of prohibitive rather than formative guidance in community rule structures.}
    \label{fig:enforcement-formation}
\end{figure*}

\section{RQ3: Alignment with Existing Surveys and Community Guidelines}

We now ask whether the cues that causally raise the odds of positive feedback are reflected in what platform users and moderators say they value and in the rules or guidelines that Reddit communities lay out for their users. We split our analysis into two parts, with \emph{RQ3(a)} comparing our causal estimates with survey-based accounts from users and moderators~\cite{weld_making_2024,lambert_positive_2024} and with descriptive inferences from highly upvoted content \cite{goyal_uncovering_2024}, and \emph{RQ3(b)} auditing existing subreddit rules and guidelines to test whether they encode these empirically supported cues to enable users to understand these \emph{``desirability norms.''}

\subsection{RQ3(a): Alignment of Causal Drivers With Prior Surveys and Descriptive Works}

\edit{A central motivation for this work was to test \emph{whether} survey-reported community values and patterns observed in highly-upvoted content actually reflect what drives positive feedback on the ground. To examine this alignment rigorously, we employed the quasi-experimental approach in our work. Below, we systematically compare our findings against prior surveys and descriptive work, revealing both where claims align with practice and where they diverge.}

\subsubsection{Comparison with prior surveys (users and moderators):} \citet{weld_making_2024} introduced a nine-category taxonomy of community values from 212 Reddit users and find that \emph{Quality of Content} is the single most frequently reported value, accounting for roughly half of all idea units, with additional emphasis on community engagement and participation. The work of \citet{lambert_positive_2024} on the other hand surveyed 115 moderators and reports that the attributes moderators want to encourage most are \emph{prosocial behavior} (46\% mention, with 24\% explicitly naming civility and respect), general \emph{quality} signals ($\approx36\%$), and \emph{participation and discussion} ($\approx23\%$).

\begin{table*}[t]
\small
\sffamily
\centering
\caption{\textbf{Policy-practice gap:} Alignment between causally-identified linguistic drivers of positive feedback (RQ1) and coverage in community guidelines across five high-performing subreddits. Community guidelines predominantly focus on basic readability and toxicity avoidance, with most empirically-supported drivers---including specificity via numbers/comparisons, future focus, and prosocial anchoring---entirely absent from written guidelines. This systematic gap between what drives positive community reception in practice and what communities explicitly teach suggests that successful linguistic norms remain largely implicit, creating barriers for newcomers and missed opportunities for formative guidance. Explicit = {\Exp}, Implicit = {\Imp}, Absent = {\Abs}.}
\label{tab:rq3b-alignment}
\Description{Table with 10 rows and 5 columns showing coverage of linguistic attributes in community guidelines across five subreddits. Three symbols indicate coverage levels: explicit checkmark circles, implicit tilde circles, and absent X circles. The first row shows explicit coverage of readability and title clarity across all five subreddits. Rows 2 through 7 covering specificity via numbers, specificity via comparisons, causal framing, future focus, light assent, and concrete social anchors show absent coverage across all subreddits, creating a uniform pattern of X symbols. Row 8 for avoiding toxicity shows mostly explicit coverage with four checkmarks and one tilde symbol. Row 9 for avoiding hedging shows absent coverage across all subreddits. Row 10 for avoiding informal tone shows mixed coverage with two explicit checkmarks and three implicit tildes. The pattern demonstrates that only 3 out of 10 empirically-supported linguistic drivers receive any coverage in community guidelines, with 7 attributes completely absent across all communities, illustrating the systematic gap between causal research findings and actual community guidance practices.}
\resizebox{0.8\linewidth}{!}{
\begin{tabular}{lccccc}
\rowcolor{gray!15}
\textbf{Linguistic Attributes} & \textbf{r/pcmasterrace} & \textbf{r/worldnews} & \textbf{r/DnD} & \textbf{r/gtaonline} & \textbf{r/unpopularopinion} \\
\cmidrule(lr){1-1}\cmidrule(lr){2-2}\cmidrule(lr){3-3}\cmidrule(lr){4-4}\cmidrule(lr){5-5}\cmidrule(lr){6-6}
Readability / title clarity      & \Exp & \Exp & \Exp & \Exp & \Exp \\
Specificity via numbers          & \Abs & \Abs & \Abs & \Abs & \Abs \\
Specificity via comparisons      & \Abs & \Abs & \Abs & \Abs & \Abs \\
Causal framing (\textit{``because,'' ``effect''}) & \Abs & \Abs & \Abs & \Abs & \Abs \\
Future focus (\textit{``will,'' ``plan''})        & \Abs & \Abs & \Abs & \Abs & \Abs \\
Light assent / prosociality      & \Abs & \Abs & \Abs & \Abs & \Abs \\
Concrete social anchors          & \Abs & \Abs & \Abs & \Abs & \Abs \\
Avoid toxicity / hostility       & \Exp & \Imp & \Exp & \Exp & \Exp \\
Avoid heavy hedging / tentative  & \Abs & \Abs & \Abs & \Abs & \Abs \\
Avoid broad informal tone        & \Exp & \Exp & \Imp & \Imp & \Imp \\
\bottomrule
\end{tabular}}
\end{table*}

Our causal results both \emph{align with} and \emph{sharpen} these survey themes. First, the calls for ``quality content'' in both surveys map closely onto multiple causal signals that improve odds: higher readability, precise and specific wording, and comparative framing. In terms of LIWC, this appears as benefits for \emph{adverbs}, \emph{comparisons}, and \emph{numbers}, along with a broadly positive \emph{cognitive residual} and \emph{causation}, all of which are consistent with clearer scaffolding, specificity, and reasoned exposition in practice. 
Second, the call for prosociality aligns with our penalties for toxicity and excessive negativity as well as small gains for light agreement-oriented cues, as we observe lower odds with negative emotion language and residual “other\_informal” category, but modest positives for \emph{assent} and \emph{family}, which can read as friendliness, and concrete social presence. Third, we find that the emphasis of the surveys on ``engagement'' and ``participation'' is only \emph{partially} supported. While we see benefits for temporal anchoring and embodied ``feel'' language, we show that stylistic patterns often used to spark discussion can backfire. In particular, the \emph{question} framing of posts and heavy connective structure are associated with \textit{lower} odds (e.g., conjunctions are disfavored; tentative language is penalized), suggesting that content that feels hedged or overly scaffolded is less rewarded even if it aims to invite dialogue. While this seems to challenges the intuition from moderator surveys that ``inciting discussion'' is generally good, we consider ``participation'' as a downstream outcome rather than a pre-feedback linguistic cause of reward, leading to this result.

At the same time, several values in the surveys fall outside what language-only, pre-feedback causal modeling can identify. For example the taxonomies include \emph{Norm adherence}, \emph{Size}, \emph{Diversity}, and \emph{Technical Features}, which are not purely textual and thus not expected to show up as causal language effects at the time of posting. 

Our findings therefore \emph{validate} the quality/prosocial cores of the surveys and \emph{refine} them into actionable linguistic levers, once author reputation, timing, and fixed effects are controlled for.

\subsubsection{Comparison with descriptive accounts:} 
\citet{goyal_uncovering_2024} analyze $16,000$ highly-upvoted comments from $80$ subreddits and, using an LLM pipeline, extract ``values'' that communities appear to encourage in practice. They report $64$ and $72$ values, extracted from $2016$ and $2022$ data, respectively. Crucially, they find that standard prosociality metrics explain only a small slice of these values, leaving about $82\%$ uncaptured.

Relative to this descriptive picture, our estimates \emph{agree} that desirability reaches well beyond generic prosociality. The strongest causal levers we identify fit within \citet{goyal_uncovering_2024}'s ``quality'' and ``accuracy/authenticity'' values, and many other effects such as \emph{focus future} and \emph{drives} fit within values such as ``futurism,'' ''progressivism,'' ``motivation,'' and ``growth mindset.'' Here, our work helps explain \emph{why} and \emph{to what extent} those values perform in practice.

Furthermore, we \emph{extend} the descriptive account by revealing that some intuitive ``upvoted'' styles are not actually causal drivers of positive feedback. For example, a lot of values extracted by \citet{goyal_uncovering_2024} did not show up as statistically significant causal predictors or even had slightly negative association in our modeling, e.g., ``historical context,'' ``philosophical,'' ``dissent.''

\subsection{RQ3(b): Auditing Community Guidelines}

We manually audit rules from five communities where our local detectors perform best (\emph{r/pcmasterrace} AUC $0.921$, \emph{r/worldnews} AUC $0.888$, \emph{r/DnD} AUC $0.864$, \emph{r/gtaonline} AUC $0.858$, \emph{r/unpopularopinion} AUC $0.848$). Given the strong performance, these communities exhibit strong, learnable linguistic regularities, making them a natural testbed for whether written guidance teaches the same cues.

We treat each rule sentence as a unit and label it for (i) \emph{function} (enforcement, formation, administrative), (ii) \emph{target} (title, body, media, meta/process), and (iii) \emph{cue alignment} with the pre–feedback levers that causally raise score in RQ1: readability/clarity, specificity via numbers, specificity via comparisons, causal framing, future focus, light assent, concrete social anchors, avoid toxicity/hostility, avoid heavy hedging/tentative, avoid broad informal tone. To measure alignment, we mark a unit \textit{explicit} if the cue is named in plain language, \textit{implicit} if hinted, and \textit{absent} otherwise.

From \autoref{tab:rq3b-alignment}, we find that the overwhelming majority of rules in all five communities regulate eligibility, formatting, and civility. On the other hand, formation guidance that teaches \emph{how to write better posts} is sparse. 
In other words, we observe a consistent \emph{``policy-practice gap'' in how community guidelines are currently written.} 
Moreover, even in cases where formation guidance exists, it focuses almost entirely on \emph{title clarity} and basic formatting. For example, \emph{r/worldnews} bans editorialized or misleading titles and all-caps words. \emph{r/gtaonline} requires descriptive titles and correct flair. \emph{r/DnD} asks for specific tags and a $400\,+$ character description for images. 

In contrast, cues that our causal models show to raise reward---\emph{numbers}, \emph{comparisons}, \emph{causal framing}, and \emph{future focus}---are virtually absent across all five sets of rules. Civility is widely encoded, which aligns with our toxicity penalties, and several communities curb broad informality in titles, which is directionally consistent with our small penalty for the \textit{informal} umbrella residual.

\subsubsection{Policy-practice gap:} Despite limited coverage of the empirically supported writing cues, our local XGBoost models for these communities achieve very high local AUCs ($0.848$ – $0.921$). This indicates strong \textit{implicit} language norms that our models can learn from behavior, even though the posted rules do not teach them explicitly. In other words, what works in practice is not fully written down in community guidelines. We also find almost no newcomer-specific guidance. None of the five rule pages explicitly advise first-time posters on how they can make their posts worthy of positive feedback, even though our analysis of \emph{RQ1(c)} shows that newcomers are disadvantaged at the baseline-level and would benefit from specific guidelines. This suggests an actionable gap for communities that want to reduce entry barriers.

As we saw, moderator and user surveys emphasize quality, civility, and participation. However, our audit shows that rules reliably encode civility and basic clarity, but they rarely articulate the pre-feedback language that causally raises the probability of the users receiving a reward, and communities appear to rely on implicit norms that participants would learn over time through observation. This gap sets up concrete opportunities to convert enforceable constraints into brief, formative guidance that reflects empirically supported writing levers.

\section{Discussion and Implications}\label{sec:discussion}

We now discuss our the key implications of our findings for designers, researchers, platforms, and moderators.

\subsection{Transforming Linguistic Patterns into Design Levers} 

Our findings reveal that linguistic choices are not merely correlates of community feedback but causal drivers that can be systematically optimized through design. This transforms language from an analytical signal into an actionable design lever that platforms can manipulate to improve community outcomes.

The most immediate application is developing real-time guidance tools that nudge users toward empirically-supported linguistic patterns. Tools like Post Guidance~\cite{horta2025post} and ConvoWizard~\cite{chang_thread_2022} could be enhanced to guide users beyond merely complying with rules toward contributing in a manner that resonate with communities. This could be done using a large language model in the form of a writing assistance tool, which can analyze and provide real-time feedback to users as they draft their contributions. These interventions should be particularly prominent for newcomers, who face baseline disadvantages ($\approx6\%$ lower odds of getting a high score) but benefit disproportionately from clarity ($65\%$ vs $40\%$ odds increase for readability). Such nudges must remain optional and easily dismissible to respect user autonomy while providing formative learning opportunities.

HCI and social computing researchers can treat specific linguistic attributes as interaction design targets, exploring interventions through simulated environments using generative agents~\cite{park2022social,anthis2025llm} before field deployment. This approach, similar to recent work on feed re-ranking~\cite{piccardi2024reranking}, would allow testing how systematically altering linguistic features affects engagement patterns and community health without risking actual user experiences.

\subsection{Early Surfacing as Socio-technical Infrastructure}

The features that affect community rewards in our causal analysis---whether statistically or not---also carry predictive signal. As we showed in \emph{RQ2}, simple XGBoost models can predict potentially desirable posts with high AUC of $0.726$ on average, before organic community upvotes accrue. Local models capture stable community regularities and a global model offers a strong default when linguistic distinctiveness is low. Used together, these detectors can help reduce the number of promising contributions that go unnoticed, while also keeping the decision-making close to community dynamics and norms.

Our call to HCI researchers and platform developers is to evaluate this capability not as a pure machine-learning achievement but as a potential for socio-technical interventions in community governance. Automated detector models for identification of undesirable content~\cite{chandrasekharan_crossmod_2019,zhan-etal-2025-slm,goyal2025momoe}, as well as systems such as ConvEx~\cite{choi_convex_2023} and Needle~\cite{liu2025needling} to aid moderators in finding problematic behavior are increasingly common. Future work should develop similar tools to surface desirable behavior (e.g., within Reddit's ``modqueue''~\cite{bajpai_modqueue_2026, lambert2025mind}) and study their effectiveness on positively reinforcing users. 

Similarly, prior work has examined ways for early detection of problematic conversations and developed tools to help proactively identify undesirable outcomes~\cite{zhang_conversations_2018,hessel_somethings_2019,chang_thread_2022,lambert_conversational_2022}.
Our predictive models could be used to complement such proactive moderation approaches and augment moderation queue sorting by prioritizing posts with high ``predicted desirability'', similar to systems contributed by prior works, such as CommentIQ~\cite{park_supporting_2016}. This could help moderators discover and amplify quality content rather than focusing solely on removal. This shift toward positive reinforcement could improve platform engagement, community health, and newcomer retention while providing formative learning about community norms.

Researchers should study deployment-time outcomes such as fairness and performance difference across communities. Our findings show that semantically distinctive communities benefit from local models while communities that have less distinctive discourse compared to other communities on average can benefit even from globally trained models. This suggests differentiated strategies across communities---some communities warrant specialized detection systems while others might adopt platform-wide standards. As a result, evaluation metrics of such deployments should include calibration across author groups and communities, not just aggregate performance.

\subsection{From Enforcement to Formation: Re-imagining Community Guidelines}

\begin{figure}
    \centering
    \includegraphics[width=\linewidth]{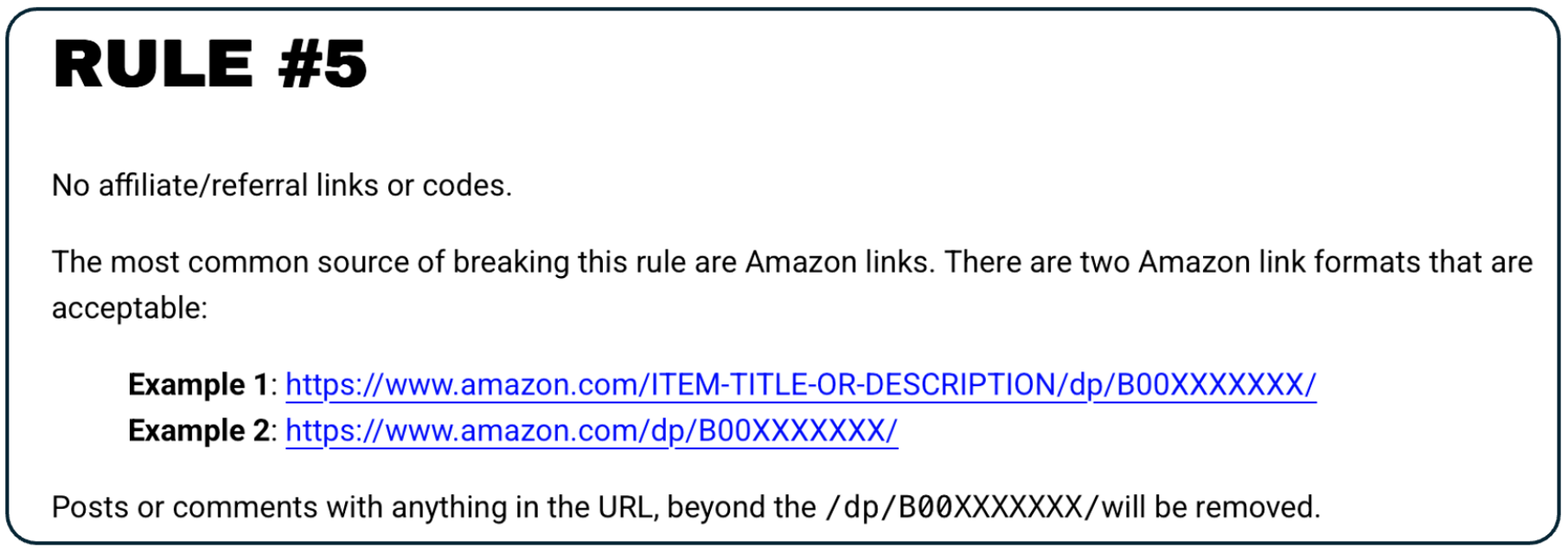}
    \caption{Detailed description of a rule from \textit{r/pcmasterrace} which teaches users how to post rule-adhering links to prevent removals by moderators. Similar exemplars for how to frame posts to align with the positive values and rewarded linguistic patterns of the community would help users make better contributions.}
    \Description{Rule documentation example showing structured community guideline format. Large header displays "RULE #5" followed by concise rule statement prohibiting affiliate and referral links. Body text explains common violations involving Amazon links and provides detailed guidance with two numbered examples showing acceptable URL formats using placeholder text. Additional enforcement details specify removal criteria for non-compliant posts. The layout demonstrates how rules can include both restrictions and specific formatting examples to help users comply, illustrating a more instructional approach to community guidelines compared to simple prohibitions.}
    \label{fig:rule-exemplar}
\end{figure} 

Our audit reveals a fundamental mismatch between what communities encode in rules and what actually drives positive feedback. While rules effectively communicate restrictions and formatting requirements, they fail to teach users strategies that increase likelihood of crafting contributions that would be positively rewarded.

This gap calls for reconceptualizing community guidelines from being purely rules for enforcement, to being formation resources. In other words, we envision that communities should use a ``capability''-focused approach in developing guidelines~\cite{jurgens-etal-2019-just} to communicate what users should be doing, rather than purely not what to do. For example, moderators could supplement restriction-heavy rules with exemplars of desirable contributions (as shown in \autoref{fig:rule-exemplar} which highlights formatting exemplar from \textit{r/pcmasterrace}) and explicit guidance on effective linguistic strategies. Platform affordances like stickied posts and community highlights could showcase high-quality contributions with annotations explaining their successful attributes. Such pedagogical approaches to facilitating cooperative behavior~\cite{grimmelmann_virtues_2015} could reduce removal rates, improve newcomer retention, and increase the overall share of well-received contributions.

System developers and researchers studying these transformation should ensure that such changes do not burden moderators. One way we hypothesize any moderator burden could be reduced is through the development of automated tools or an approach utilizing a large language model which could analyze existing high-performing posts to generate community-specific writing tips, creating a feedback loop where successful patterns become exemplars for future contributors.

\subsection{Advancing Theory, Data, and Methods in Online Community Research}

This work makes several theoretical and methodological contributions that extend beyond Reddit. We demonstrate that positive reinforcement operates through measurable, causal linguistic mechanisms rather than abstract notions of ``quality,'' providing a framework for understanding how textual choices translate into community reception. Our stratified causal approach isolates linguistic effects from confounds like reputation and timing, offering a template for studying other platform behaviors where selection effects typically obscure causality.

Our work also contributes a comprehensive feature set spanning 100+ linguistic, semantic, and psychosocial attributes that future researchers can directly apply or augment for studying online feedback mechanisms. This feature collection captures multiple facets of textual communication that influence community reception. Researchers can leverage this curated feature set as a foundation for studying feedback dynamics on other platforms, investigating cross-cultural variations, or exploring adjacent phenomena like virality~\cite{chan_understanding_2024}, engagement~\cite{chan2025examining}, or community norm formation~\cite{chandrasekharan_internets_2018, reddy_evolution_2023}. \edit{We hypothesize that while core linguistic features (clarity, toxicity) would likely generalize across platforms, Reddit-specific affordances (threaded discussions, voting) may limit transferability to platforms with different structures.}

Future research should conduct field experiments to test whether implementing these linguistic strategies improves longitudinal outcomes. Researchers should also explore cross-platform generalization, and explore whether there are cultural and linguistic variations in non-English communities, and study how deployment of such interventions might shift community linguistic norms over time. Platform ranking audits~\cite{piccardi2024reranking,chan2025examining} could also interleave detector-ranked content with organic ranking to measure long-term effects on community dynamics, author diversity, and norm evolution.

\subsection{Ethical Considerations: Community Rewards $!=$ Social Values}

Our findings reveal which linguistic patterns drive positive community feedback. However, what communities reward may diverge from what is most valuable, either to individual users or to broader social outcomes~\cite{diakopoulos_towards_2011}. We explicitly note that upvotes represent \emph{community approval rather than a guarantee of accuracy or quality} (Section 3.1), yet communities may systematically reward entertaining over informative content, consensus-reinforcing over challenging perspectives, and discussion-generating over problem-solving contributions. This is not a methodological limitation but rather a fundamental feature of crowd-based feedback systems.

This divergence could create equity concerns. For example, we show question-heavy, help-seeking posts face lower odds of high scores, while newcomers are already disadvantaged. A user genuinely seeking technical help may need to write precisely the style our analysis shows communities penalize. An important consideration here is that optimizing for upvotes $!=$ optimizing for utility to the poster. Therefore, we argue that guidance systems should be goal-aware (asking users whether they want visibility, answers, or discussion), optional and dismissible to preserve autonomy, and transparent about trade-offs. Rather than teaching universal reward optimization, platforms should consider algorithmic interventions to surface diverse contribution types, protect help-seeking behavior, and provide newcomer affordances. Our findings are a necessary input to building better systems, but the path from empirical findings to ethical design requires careful consideration of whose interests are served and how to preserve authenticity of participation.

\subsection{Limitations and Future Work}

Our work has limitations that point to promising directions for future research.

\subsubsection{Temporal Generalizability:} Our analysis focuses on a specific five-month window, and the linguistic drivers of positive feedback we identify may not generalize across different time periods, particularly as platform norms, user demographics, and cultural contexts evolve. Future work should investigate the temporal stability of these linguistic patterns by analyzing data across multiple years and examining how reward mechanisms shift over time. 

\subsubsection{Evaluation Gap:} While our models achieve strong predictive performance, we do not validate these predictions against human judgments of post quality. Future research should conduct controlled human evaluation studies where moderators and users rank pairs of posts, comparing these rankings against our model predictions. Such validation would help determine whether our approach captures genuine quality signals or merely reflects voting patterns that may diverge from human perceptions of valuable contributions. Further, we do not evaluate the potential of transformer-based or large language models on this prediction task~\cite{goyal2026vastu}.

\edit{\subsubsection{Sensitivity to Unmeasured Confounding:} Our causal claims are conditional on the assumption that residual linguistic variation is as good as random after controlling for observed covariates. We conducted sensitivity analyses (\autoref{app:sensitivity-analysis}) which suggest that unmeasured confounding would need to be implausibly strong with a partial $R^2 > 0.15$, exceeding our strongest observed confounder at $R^2 \approx 0.12$) to eliminate key effects, though more fragile findings may be more vulnerable to moderate confounding. Future work  could address this through field experiments that randomize linguistic guidance  or natural experiments that provide exogenous variation in writing style, but  our quasi-experimental approach represents the most rigorous feasible design for studying naturally occurring linguistic choices at scale.}

\edit{\subsubsection{Visual Content:} Our analysis focuses exclusively on textual features and cannot account for  visual content (images, videos, memes), which likely plays a substantial role in community feedback, particularly in visually-oriented subreddits where our  text-based models underperform.}

\subsubsection{Cross-Community Transferability:} Our analysis treats each community independently, but we do not systematically evaluate how predictive models trained on one community perform when applied to others. Future work should conduct comprehensive cross-community or even cross-platform evaluation to assess model transferability, identify clusters of communities with similar linguistic reward patterns, and develop approaches that leverage shared signals while maintaining sensitivity to local norms.

\subsubsection{Limited Scope of Guidelines Audit:} Our audit of community guidelines covers only five high-performing communities and focuses specifically on the linguistic attributes we identify as causal drivers. A more comprehensive understanding of the policy-practice gap would require large-scale analysis across hundreds of communities, examining a broader range of guidance types and community governance approaches. Future research should develop automated methods for analyzing community guidelines at scale, investigating how rules varies across communities, and studying the relationship between explicit guidance and implicit behavioral norms across diverse online spaces.

\section{Conclusion}

Through an empirical study of posts from 100 subreddits, we show that the language that authors use in their posts causally affects the rewards and reception they receive from the community. Using risk stratification and fixed effects based causal modeling, we identify linguistic attributes that influence the odds of positive reception at the moment of posting, and we demonstrate that these same attributes can be used to support early surfacing by training predictive models. We contextualize these results in prior work, finding both overlap and tension with prior survey and descriptive accounts. Finally, we audit community rules to reveal a persistent gap between what is taught and explicitly stated in community guidelines, and what works in practice. Together, these findings suggest a path for human-centered systems that teach and surface desirable contributions to users. Future work should evaluate these interventions through field experiments, report who benefits, track calibration over time and across communities, and partner with moderators and users to ensure that guidance remains minimally intrusive and aligned with local norms of the community.

\begin{acks}
This research was supported by NSF CAREER IIS-2439433. 

A.G. was supported by compute credits from the OpenAI Researcher Access Program. Furthermore, this work used the Delta system at the National Center for Supercomputing Applications through allocation \#240481 from the Advanced Cyberinfrastructure Coordination Ecosystem: Services \& Support (ACCESS) program, which is supported by National Science Foundation grants \#2138259, \#2138286, \#2138307, \#2137603, and \#2138296.

We thank the members of the Social Computing Lab (SCUBA), especially Jackie Chan, for their feedback and input on this work. We also thank Yiren Liu and Yijun Liu for helpful suggestions.
\end{acks}

\bibliographystyle{ACM-Reference-Format}
\bibliography{references}

\newpage

\appendix

\section{LLM Prompt for Interpreting Principal Components}\label{app:llm-prompt}

\begin{figure}
    \centering
    \includegraphics[width=0.95\linewidth]{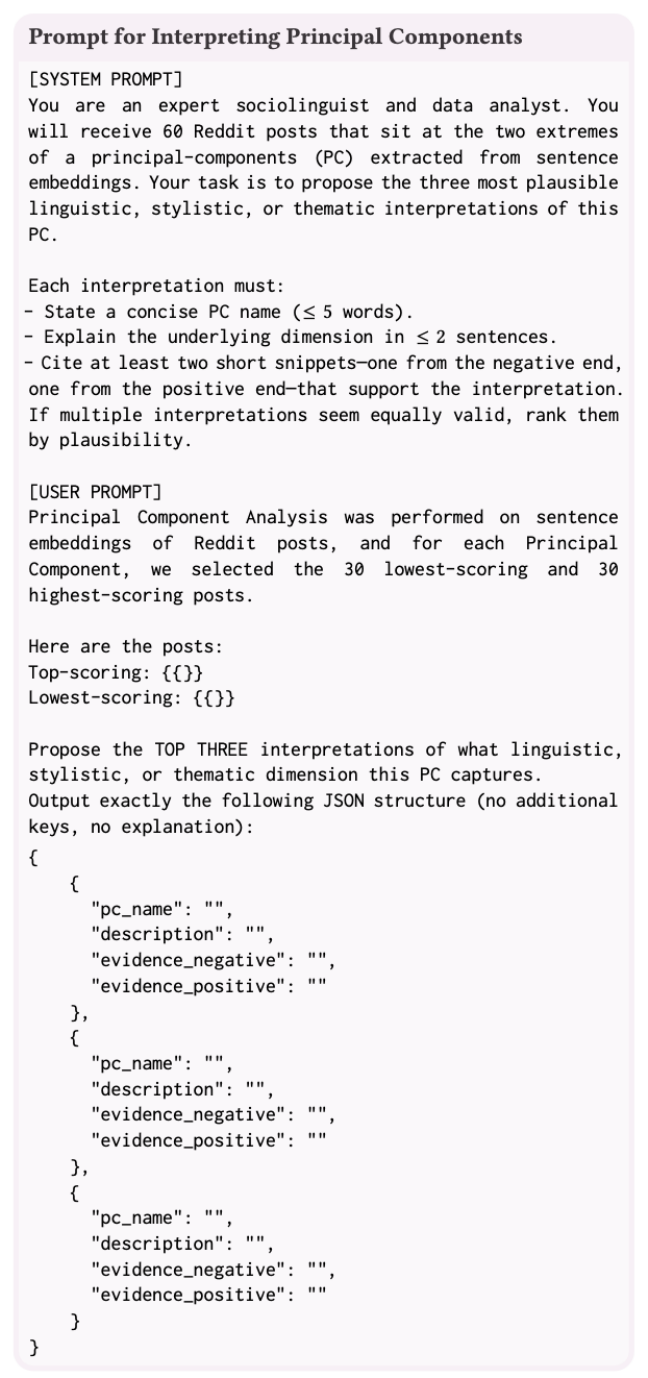}
    \caption{Prompt for Principal Component Interpretation}
    \Description{Structured large language model prompt displayed in a bordered text box. The prompt contains two main sections: a system prompt establishing the role of expert sociolinguist analyzing Reddit posts at extremes of principal components, and a user prompt requesting three interpretations with specific formatting requirements. Each interpretation must include a concise name under five words, a two-sentence explanation, and supporting evidence from both positive and negative ends. The bottom portion shows a JSON template structure with four fields per interpretation: pc_name, description, evidence_negative, and evidence_positive. The prompt specifies exact output formatting requirements and includes placeholders for the actual Reddit post data to be analyzed.}
    \label{fig:prompt-pca}
\end{figure}

\autoref{fig:prompt-pca} shows the LLM prompt used for interpreting the extracted principal components.

\edit{\section{Temporal Robustness Check}
\label{app:temporal-robustness}}

\edit{To assess whether our findings are stable across time or reflect period-specific effects, we split our observation window into two non-overlapping periods and re-estimated all models independently:}
\edit{\begin{itemize}
    \item \textbf{Window 1:} May 1 - July 14, 2020 (\emph{Baseline:} May 1-14; \emph{Sampling:} May 15 - July 14)
    \item \textbf{Window 2:} July 15 - September 30, 2020 (\emph{Baseline:} July 15-28; \emph{Sampling:} July 29 - September 30)
\end{itemize}}
\edit{For each window, we applied the identical methodology described in Section 3. Specifically, we constructed candidate pools for high score, computed baseline covariates from the 14-day baseline periods, performed risk stratification, and estimated logistic regression models with fixed effects. We then extracted the log-odds ratios for all significant features ($q < 0.05$ after BH-FDR correction) from each window and computed the Pearson correlation between the two sets of estimates.}

\edit{The correlation between log-odds ratios across the two time periods is $r = 0.764 ~(p < 0.001)$, indicating substantial consistency in which linguistic features drive positive feedback and the magnitude of their effects. This suggests our findings are not artifacts of a specific moment but reflect more stable patterns in how Reddit communities respond to linguistic variation. While some variation exists likely due to different sample compositions and period-specific events—the core drivers (readability, discussion-generating framing, question-avoidance, toxicity penalties) remain consistent across both time windows.}

\edit{\section{Sensitivity Analysis to Unobserved Confounders}\label{app:sensitivity-analysis}}

\edit{Our causal estimates assume that conditional on observed covariates, residual linguistic variation is as good as random with respect to outcomes. While we built in extensive control for confounders, unmeasured factors could bias results so we assess how strong such confounding would need to be to overturn findings.}

\edit{We compute partial $R^2$ (variance explained after accounting for other variables) for our strongest observed confounders as a baseline:
\begin{itemize}
    \item Daily posting rate: $R^2 = 0.08$
    \item Daily removal rate: $R^2 = 0.03$
    \item Average score on the author's posts: $R^2 = 0.12$
    \item Author's Reddit account age: $R^2 = 0.06$
\end{itemize}
These numbers indicate how influential unmeasured variables would need to be to substantially bias results.}

\edit{Since our setting has 100+ continuous treatments rather than one binary treatment, multiple fixed effects, and risk stratification that standard sensitivity packages don't accommodate, instead of using a traditional Rosenbaum-styled sensitivity check~\cite{rosenbaum2004design} we simulate unmeasured confounders of varying strength for each key feature and add them to our model to observe when the estimated effect becomes non-significant. In other words, this analysis approximates how strong an unmeasured confounder, measured by its partial $R^2$ with both treatment and outcome would need to be in order to eliminate the observed effect.}

\edit{We focus on key features of \emph{readability}, \emph{discussion framing}, \emph{question ratio}, and \emph{toxicity}. We find that in order to reduce effects to null (OR = 1.00) for \emph{Readability} (OR=1.40), we require $R^2 \ge 0.15$, for \emph{Discussion framing} (OR=1.43), we require $R^2 \ge 0.16$, for \emph{Question ratio} (OR=0.71), we require $R^2 \ge 0.13$, and for \emph{Toxicity} (OR=0.95), we require $R^2 \ge 0.02.$}

\edit{Therefore, for these effects (except toxicity), unmeasured confounding would need $R^2 > 0.13$, which is stronger than our most powerful observed confounder with $R^2 = 0.12$. We believe that other plausible confounders such as author mood, external events, topic expertise, etc. are unlikely to be this strong, especially given our extensive controls. However, the toxicity effect is relatively more fragile, requiring only $R^2 \ge 0.02$ to eliminate effect.}

\end{document}